\documentclass[fleqn,usenatbib]{mnras}
\usepackage[T1]{fontenc}
\usepackage{ae,aecompl}
\usepackage{setspace}
\usepackage{natbib}
\usepackage{textcomp}
\usepackage{amsmath,amssymb,amsfonts,fullpage}
\usepackage{txfonts}
\usepackage{subfigure}
\usepackage[dvips]{graphicx}

	\title{\bf{Moonfalls: Collisions between the Earth and its past moons}}

	\author
	{		
	Uri Malamud,$^{1}$
	Hagai B. Perets,$^{1}$
	Christoph Sch{\"a}fer,$^{2}$
	Christoph Burger$^{3}$
	\\	
	$^{1}$Department of Physics, Technion Israeli Intitute of Technology, Technion City, 3200003 Haifa, Israel\\
	$^{2}$Institut f{\"u}r Astronomie und Astrophysik, Eberhard Karls Universit{\"a}t T{\"u}bingen, Auf der Morgenstelle 10, 72076 T{\"u}bingen, Germany\\
	$^{3}$Department of Astrophysics, University of Vienna, T{\"u}rkenschanzstra{\ss}e 17, 1180 Vienna, Austria\\
	}

	\pubyear{2018}
	
\begin{document}
		
	\label{firstpage}
	\pagerange{\pageref{firstpage}\textendash{}\pageref{lastpage}}
	\maketitle
	
	\begin{abstract}
	
	During the last stages of the terrestrial planet formation, planets
	grow mainly through giant-impacts with large planetary embryos. The Earth's Moon was suggested to form through one of these impacts. However, since the proto-Earth has experienced many giant-impacts, several moons (and also the final Moon) are naturally expected to form through/as-part-of a sequence of multiple (including smaller scale) impacts. Each impact potentially forms a sub-Lunar mass moonlet that interacts gravitationally with the proto-Earth and possibly with previously-formed moonlets. Such interactions result in either moonlet-moonlet mergers, moonlet ejections or infall of moonlets on the Earth. The latter possibility, leading to low-velocity moonlet-Earth collisions is explored here for the first time. We make use of smooth particle hydrodynamical (SPH) simulations and consider a range of moonlet masses, collision impact-angles and initial proto-Earth rotation rates. We find that grazing/tidal-collisions are the most frequent and produce comparable fractions of accreted-material and debris. The latter typically clump in smaller moonlets that can potentially later interact with other moonlets. Other collision geometries are more rare. Head-on collisions do not produce much debris and are effectively perfect mergers. Intermediate impact angles result in debris mass-fractions in the range of 2-25\% where most of the material is unbound. Retrograde collisions produce more debris than prograde collisions, whose fractions depend on the proto-Earth initial rotation rate. Moonfalls can slightly change the rotation-rate of the proto-Earth. Accreted moonfall material is highly localized, potentially explaining the isotopic heterogeneities in highly siderophile elements in terrestrial rocks, and possibly forming primordial super-continent topographic features. Our results can be used for simple approximations and scaling laws and applied to n-body studies of the formation of the Earth and Moon. 
	
	\end{abstract}
	

	\section{Introduction}\label{S:Intro}
	The multiple-impact origin of the Earth's Moon represents an extended and more comprehensive apprehension of the Moon's formation, compared with the current paradigm of a single giant-impact origin of the Moon. Simulations show \citep{Canup-2004,Canup-2014} that the single giant-impact hypothesis can reproduce, under certain assumptions,
	many of the present-day characteristics of the Earth-Moon system.
	It is however also challenged by several difficulties, including the
	isotopic composition similarity problem (\citealt{PahlevanStevenson-2007,ZhangEtAl-2012,Asphaug-2014,HerwartzEtAl-2014,KaibCowan-2015,KruijerEtAl-2015},
	but see \citealt{Mastrobuono-BattistiEtAl-2015,Mastrobuono-BattistiPerets-2017}), the mutual inclination problem
	\citep{ToumaWisdom-1998,WardCanup-2000,PahlevanMorbidelli-2015,CukEtAl-2016}, the low probability of obtaining the canonical Earth-Moon forming giant impact \citep{RaymondEtAl-2009,Mastrobuono-BattistiPerets-2017,RufuEtAl-2017} and the requirement to circumvent full homogenization of the Earth's mantle
	\citep{NakajimaStevenson-2015,PietEtAl-2017}. In addition, it is
	intrinsically incomplete, since only the last giant impact is considered, while the critical collisional evolution taking place during planetary accretion \citep{AgnorEtAl-1999,Chambers-2001,CitronEtAl-2014,JacobsonMorbidelli-2014,KaibCowan-2015,Mastrobuono-BattistiEtAl-2015,Mastrobuono-BattistiPerets-2017,RufuEtAl-2017} prior to and possibly following this single event is ignored.
	
	The multiple-impact origin hypothesis is rooted in a more realistic view of terrestrial planet formation. It envisions multiple planetary-scale impacts resulting in the late stages of planet formation \citep{Ringwood-1989,CitronEtAl-2014,RufuEtAl-2017,CitronEtAl-2018}. Each of these impacts could have formed a debris disc that in turn typically coagulates to form a single moon \citep{IdaEtAl-1997,KokuboEtAl-2000,SalmonCanup-2012,SasakiHosono-2018}. In less probable cases where it forms an additional large moon, \citet{CanupEtAl-1999} show that this configuration will be unstable and will evolve to form a single moon as one moon merges with the other or falls to the Earth. Assuming a single coagulated moon, multiple impacts will subsequently produce a multi-moon system anyway. The tidal evolution and migration of these formed moons and their mutual gravitational perturbations \citep{CitronEtAl-2018} could then result in several possible evolutionary outcomes, including collisions among two moons, ejection of moons, or their re-collisions with the proto-Earth (either from moon-moon interactions or de-orbiting of retrograde moons \citep{Counselman-1973}). Within the framework of the multi-impact origin, the specific purpose of this paper is to study the implications of collisions between the proto-Earth and infalling moons.
	
	Through the use of hydrodynamical codes, past studies have already examined the implications of collisions between the proto-Earth and planetary-embryo-sized impactors \citep{CanupAsphaug-2001,Canup-2004,WadaEtAl-2006,CanupEtAl-2013,Canup-2014,HosonoEtAl-2016}, in addition to smaller impactors, roughly in the lunar-to-planetary-embryo mass range \citep{RufuEtAl-2017,MarchiEtAl-2017}. In the general context of terrestrial planetary accretion and formation, the velocities of such impactors relative to the mutual escape velocity were varied, and in the range of approximately $1<v/v_{esc}<4$. No other previous study, with the exception of \citet{MarchiEtAl-2017} has however investigated the collision between the proto-Earth and sub-Lunar-mass impactors. In particular, infalling impactors, i.e., impactors originating from a bound system of multiple moons, have impact velocities strictly of the order of $v/v_{esc}=1$. Such low velocities were not explored by \citet{MarchiEtAl-2017}. In this study, we use a hydrodynamical code in order to examine such collisions for the first time.
	
	Our specific goals are to assess and study the parameter space associated with impactor mass and impact geometry, and examine their effect on the particular composition as well as origin of material, either accreting onto the proto-Earth, remaining in bound orbit around it or escaping the system entirely. We also evaluate how such collisions affect the proto-Earth's rotation rate and the distribution of accreted impactor material. This paper serves to examine one aspect in a larger research plan, whose goal is to connect different aspects of the multiple-impact hypothesis. In what follows we briefly introduce in Section \ref{S:Methods} the model used for hydrodynamical simulations, the considerations for our parameter space, and introduce our pre- and post-processing algorithms. In Section \ref{S:Results} we present the simulation results, and discuss their implications in the framework of the multiple-impact origin hypothesis in Section \ref{S:discussion}.

	\section{Methods}\label{S:Methods}
	
	\subsection{Hydrodynamical code outline}\label{SS:Outline}
	We perform hydrodynamical collision simulations using an SPH code developed by \citet{SchaferEtAl-2016}. The code is implemented via CUDA, and runs on graphics processing units (GPU), with a substantial improvement in computation time, on the order of several $\sim10^{1}-10^{2}$ times faster for a single GPU compared to a single CPU, depending on the precise GPU architecture. While having been developed rather recently (based on an earlier OpenMP implementation), the code has already been successfully applied to several studies involving impact processes \citep{DvorakEtAl-2015,MaindlEtAl-2015,HaghighipourEtAl-2016,BurgerEtAl-2018}.
	
	The code implements a Barnes-Hut tree that allows for treatment of self-gravity, as well as gas, fluid, elastic, and plastic solid bodies and also includes a failure model for brittle materials. Given the analysis of \citet{BurgerSchafer-2017} however, and the typical mass and velocity considered for our impactors and target (see Section \ref{SS:ParameterSpace}), we perform our simulations in full hydrodynamic mode, i.e., neglecting solid-body physics, being less computationally expensive.
	
	For relating the material pressure $P$, density $\rho$, and specific internal energy $e$, we use the Tillotson equation of state (EOS). The parameters for the EOS are taken from \citet{Melosh-1989} and \citet{BenzAsphaug-1999}, for iron and silicate (basalt) respectively. The Tillotson EOS is relatively simple to implement but still applicable over a wide range of physical conditions. It has two distinct analytical forms, given according to $\rho$ and $e$ relative to the zero pressure density $\rho_{0}$, the energy of incipient vaporization $e_{iv}$ and the energy of complete vaporization $e_{cv}$ (see e.g. \citet{BurgerEtAl-2018}) which determine the relevant parameter domains. These domains are essentially a crude estimation of the thermodynamical state, and therefore despite the many advantages of the Tillotson EOS, results based on it should be regarded as merely an approximation. While ANEOS, a more complex and thermodynamically consistent EOS developed by the Sandia National Laboratory, is also implemented in the code, it is not yet fully and sufficiently tested. We nevertheless point to a recent analysis by \citet{EmsenhuberEtAl-2018} which demonstrates that the Tillotson EOS produces comparable results to ANEOS, particularly in collision regimes where vaporization is not of crucial importance. Given the results presented in Section \ref{SS:Localization}, in most cases the use of the Tillotson EOS is justified here.
	
	Since the hydrocode used in this paper is relatively new, we performed a set of preliminary verification tests. We note that this is in addition to the work of \citet{SchaferEtAl-2016}, which also performed several successful verification tests, including colliding elastic rings, the gravitational collapse of a molecular cloud, impact cratering and planetary-scale collisions. Our own test consisted of a Moon forming, giant impact scenario, which was chosen to match the impact variables (impactor/target mass ratio, velocity, structure and impact angle) given by \citet{Canup-2004} and later repeated in more detail by \citet{CanupEtAl-2013}, as portrayed in their Figures 2 and 4 respectively. Our comparison using $5 \cdot 10^{4}$ as well as $10^{5}$ SPH particles, yielded qualitatively identical results. A quantitative comparison between our test results and those by \citet{CanupEtAl-2013} proved to be inaccurate due to differences in the EOS used in both studies and also the use of different criteria for estimating disc mass (see Section \ref{SS:Algorithms} for our mass estimation algorithm versus their Section 3.4). However when we adapted our criteria to match theirs, our results were in close agreement: our obtained disk mass as a function of time was within 10$\%$ of theirs and our final disk impactor mass fraction was within 2-3$\%$ of theirs. These deviations are typically smaller than the ones they received when comparing different resolutions and methods, albeit using the exact same code. Following these qualitative and quantitative analyses, we are sufficiently satisfied that our code compares well against one of the most widely and extensively used codes in the literature, a descendant of the seminal code by \citet{BenzEtAl-1989}. 
	
	\subsection{Parameter space}\label{SS:ParameterSpace}
	Our collision scenarios consist of bound and therefore low velocity impactors of various masses impacting onto an Earth-mass planet. For simplicity we assume a uniform impactor velocity that equals the mutual escape velocity $v_{esc}$. We analyse the data from the \citep{CitronEtAl-2018} study and find a distribution of impact velocities relevant to infalling moonlets. According to this distribution $v_{esc}$ is a reasonable and judicious assumption and may be regarded as an upper limit. Both the impactor and target are differentiated with an iron core and silicate mantle, whose mass fractions are similar to those of the present day Earth and Moon, respectively. For the proto-Earth we thus have a core whose mass fraction is $33\%$ underlying a mantle whose mass fraction is $67\%$. Although the iron core of the Moon is merely of the order of $\sim2\%$ in mass \citep{Wieczorek-2006}, thereby having only a negligible contribution in our simulations, we still account for it, as it allows us to infer the qualitative behaviour of more substantial impactor iron cores.
	
\begin{figure}
	\begin{center}	
		\includegraphics[scale=0.245]{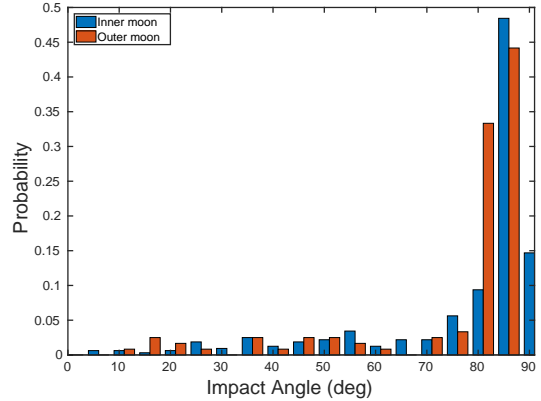}
		\caption{The distrbution of moonlet-Earth impact angles (0\textdegree - head-on impact; 90\textdegree - exteremely grazing impact) based on analysis of data from the \citet{CitronEtAl-2018} study. As can be seen most collisions are extremely grazing, but the overall distribution includes a wide range of rarer head-on and intermediate impact angles.}
		\label{fig:impact-angles}
	\end{center}	
	\end{figure}

	The impactors are modelled to have sub-Lunar masses $M_{imp}$ of 1/2, 1/3 and 1/6 the mass of the Moon ($M_{L}$). We further consider three impact parameters: (a) near head-on, (b) intermediate and (c) extremely grazing, where $b= \sin \theta$ is the impact parameter, and $\theta=\{22$\textdegree$,51$\textdegree$,89$\textdegree$\}$ the impact angle, that is, the angle between the impactor's trajectory and the target's surface normal at the time of first contact. In Figure
	\ref{fig:impact-angles} we show the distribution of impact angles
	based on analysing the results from N-body simulations of moonlet-moonlet interactions in the \citet{CitronEtAl-2018} study. As can be seen most impacts are extremely grazing, but the overall distribution includes a wide range of more rare head-on and intermediate impacts, which is why we consider all three possibilities. Likewise, retrograde tidally de-orbiting moonlets result in highly grazing impacts as well. We thus have $b=\{0.375$, $0.777$ and $0.999\}$, respectively. Note that here we consider only impacts that occur in the proto-Earth's equatorial plane.
	
	The last free parameter is the proto-Earth's pre-collision rotation rate. We consider $\omega$ = 0, 0.1, 0.25 and 0.5 $\omega_{max}$, where $\omega_{max}$ is the planetary rotation rate at breakup, and corresponds to a 1.475 h rotation period. This choice is taken to be compatible with \citet{RufuEtAl-2017}. For $\omega>0$, we consider both prograde and retrograde impacts. We note that while $\omega=0$ is inconsistent with the assumption of moonlet tidal migration (and subsequent gravitational perturbations leading to infall), we still consider it as a viable (though unlikely) possibility, since an additional accretion event can theoretically terminate the proto-Earth's rotation during migration and prior to infall.
	
	Altogether we have $3\times3=9$ simulation cases with $\omega=0$, each having $10^{5}$ SPH particles. We also performed 9 additional simulations with $\omega=0$ and $5 \cdot 10^{4}$ SPH particles, in order to study the effect of reducing SPH resolution by a factor of two. Since in Section \ref{SS:Time} we show the low resolution results to be saisfactory, the rest of the $3 \times 3 \times 6 =54$ simulations that have $\omega>0$ are carried out in lower resolution given the significant saving in computation time. Together with the test runs we have a total of 74 simulation cases. The test simulations and $\omega=0$ simulations were performed on the 'TAMNUN' GPU cluster, at the Technion Institute in Israel. The GPU model used is NVIDIA Tesla K20. Each high resolution simulation ran on a single dedicated GPU for approximately 4-5 weeks, tracking up to the first 84 hours post collision. Low resolution $\omega>0$ simulations were performed on the bwForCluster BinAC, at T{\"u}bingen University in Germany. The GPU model used is NVIDIA Tesla K80. Each simulation ran on a single dedicated GPU for several days up to one week, tracking the first 55 hours post collision. In total we thus have $\sim$100 weeks of GPU time.
	
	\subsection{Pre and Post processing algorithms}\label{SS:Algorithms}
	The initial setup of each simulated scenario with $\omega=0$ (i.e., no initial target rotation) is calculated via a pre-processing step, in which both impactor and target are generated with relaxed internal structures, i.e. having hydrostatic density profiles and internal energy values from adiabatic compression, following the algorithm provided in appendix A of \citet{BurgerEtAl-2018}. This self-consistent semi-analytical calculation (i.e., using the same constituent physical relations as in the SPH model) equivalently replaces the otherwise necessary and far slower step of simulating each body in isolation for several hours, letting its particles settle into a hydrostatically equilibrated state prior to the collision (as done e.g., in the work of \citet{CanupEtAl-2013} or \citet{SchaferEtAl-2016}). The initial configuration then places the colliding bodies at a distance of a few times the sum of their radii. The two bodies are given velocities according to the desired collision parameters mentioned in Section \ref{SS:ParameterSpace}.
	
	For simulated scenarios with $\omega>0$ (i.e., with initial target rotation) a slightly different procedure is employed. Since the relaxation code of \citet{BurgerEtAl-2018} does not account for rotation, an additional relaxation phase is required in order to damp any initial radial oscillations in the target. We therefore start out by initializing the internal structure of target and impactor using the same technique as in the previous paragraph. However the binary is configured such that the impactor and target are placed wide apart, at a distance of $\sim$40 times the sum of their radii. The rotating target then has sufficient time in order to settle into a stable rotation prior to the impact.
	
	After the SPH simulation is concluded, outputs of SPH particles and their inherent properties, including their densities, velocities, energies etc., are analysed. From the aforementioned particles we derive bulk quantities, such as the eventual mass of particles that will accrete onto the proto-Earth, remain in a bound disc, or escape the system entirely. We also wish to track the precise composition and origin of the particles in their respective end destinations, as well as to calculate the proto-Earth's change in rotation rate (magnitude and rotation axis vector). All of these bulk quantities are time dependent, and must be calculated via a post-processing step, as follows.
	
	In order to estimate the final classification of particles in the simulation we apply the following 5-step algorithm.
	
	(a) We find physical fragments (clumps) of spatially connected SPH particles using a friends-of-friends algorithm. 
	
	(b) The fragments are then sorted in descending order.
	
	(c) We classify these fragments in two categories: gravitationally bound (GB) to the proto-Earth and gravitationally unbound (GUB). The first fragment (i.e., the most massive, in this case the target/proto-Earth) is initially the only one marked as GB, and the rest are marked as GUB. 
	
	(d) We calculate: 
	\begin{equation}
	\vec{r}_{GB}=\frac{\sum_{j}m_{j}\vec{r}_{j}}{\sum_{j}m_{j}},\vec{v}_{GB}=\frac{\sum_{j}m_{j}\vec{v}_{j}}{\sum_{j}m_{j}}
	\label{eq:CenterOfMass}
	\end{equation}
	where $\vec{r}_{GB}$ and $\vec{v}_{GB}$ are the center of mass position and velocity of GB fragments, $j$ denoting indices of GB fragments and $m_{j}$, $\vec{r}_{j}$ and $\vec{v}_{j}$ are the corresponding mass, position and velocity of each fragment.
	
	Then, for each fragment marked as GUB we check if the kinetic energy is lower than the potential energy: 
	\begin{equation}
	\frac{\lvert\vec{v}_{i}-\vec{v}_{GB}\rvert^{2}}{2}<\frac{G\left(M_{GB}+m_{i}\right)}{\lvert\vec{r}_{i}-\vec{r}_{GB}\rvert}
	\label{eq:Bound}
	\end{equation}
	where $G$ is the gravitational constant and $M_{GB}$ is the summed mass of gravitationally bound fragments. $m_{i}$, $\vec{r}_{i}$ and $\vec{v}_{i}$ are the fragment mass, position and velocity, $i$ denoting indices of GUB fragments. If equation \ref{eq:Bound} is satisfied then fragment $i$ is switched from GUB to GB. We iterate on step (d) until convergence (no change in fragment classification within an iteration) is achieved.
	
	(e) We now calculate the respective fragment energy $E_{j}$, eccentricity $e_{j}$ and pericentre $q_{j}$, $j$ denoting the indices of GB fragments that were identified following the previous step: 
	\begin{equation}
	E_{j}=\frac{m_{j}{\lvert\vec{v}_{j}-\vec{v}_{GB}\rvert}^{2}}{2}-\frac{Gm_{j}M_{GB}}{\lvert\vec{r}_{j}-\vec{r}_{GB}\rvert}
	\label{eq:Energy}
	\end{equation}
	\begin{equation}
	e_{j}=\sqrt{1+\frac{2E_{j}{\lvert\vec{v}_{j}-\vec{v}_{GB}\rvert}^{2}{\lvert\vec{r}_{j}-\vec{r}_{GB}\rvert}^{2}}{G^{2}M_{GB}^{2}m_{j}}}
	\label{eq:Eccentricity}
	\end{equation}
	\begin{equation}
	q_{j}=\frac{{\lvert\vec{v}_{j}-\vec{v}_{GB}\rvert}^{2}{\lvert\vec{r}_{j}-\vec{r}_{GB}\rvert}^{2}\sqrt{1-e_{j}^{2}}\left(1-e_{j}\right)}{GM_{GB}}
	\label{eq:Pericentre}
	\end{equation}
	
	Each fragment that satisfies $q_{j}>2.9R_{P}$, where $R_{P}$ is the proto-Earth's radius, has its pericentre beyond the fluid Roche limit and may be finally classified as belonging to the bound debris disc, otherwise it is classified as part of the proto-Earth.
	
	Once the algorithm is complete we have $M_{U}$, $M_{P}$ and $M_{D}$, the summed masses of unbound, proto-Earth and disc fragments, respectively. We also track for each one, their respective compositions and the origin of SPH particles (impactor or target).
	
	The change in target rotation rate (magnitude and axis) is calculated as follows. For the largest fragment of spatially connected SPH particles, we find the total angular momentum $\vec{L}=\sum m_{p}\left(\vec{r}_{p}\times\vec{v}_{p}\right)$ by the summation of its individual particle angular momenta, where $m_{p}$ denotes SPH particle mass and $\vec{r}_{p}$ and $\vec{v}_{p}$ denote SPH particle relative (to fragment's center of mass) position and velocity. We then calculate the angular momentum unit vector $\hat{L}=\vec{L}/\lvert\vec{L}\rvert$, which points towards the direction of the rotation axis. In order to get the rotation time $P_{rot}$ we calculate $\vec{R}=\hat{L}\times\vec{r}_{p}$, the distance vector from the rotation axis to the particle relative position. Then the total moment of inertia is similarly given by $I=\sum m_{p}{\lvert\vec{R}\rvert}^{2}$, and the rotation rate $P_{rot}=2\pi I/\lvert\vec{L}\rvert$.

	\section{Results}\label{S:Results}
	
	\subsection{Resolution dependence and temporal analyses}\label{SS:Time}
	\label{SS:Time} 
\begin{figure}
		\centering{}\includegraphics[scale=0.53]{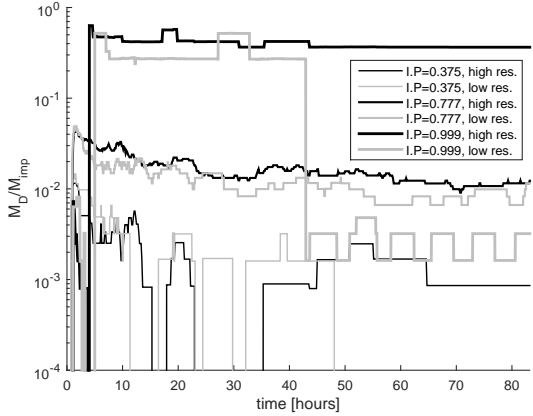} \caption{Bound disc mass as a function of time, given various impact parameters and SPH resolutions. For a 1/2 $M_{L}$ impactor mass.}
		\label{fig:TimeDiscMass} 
	\end{figure}

\begin{figure}
		\centering{}\includegraphics[scale=0.53]{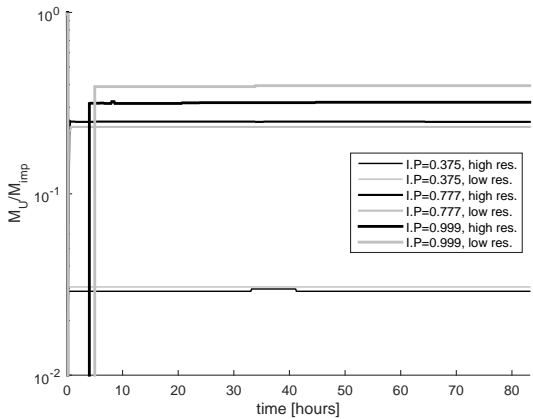} \caption{Unbound material mass as a function of time, given various impact parameters and SPH resolutions. For a 1/2 $M_{L}$ impactor mass.}
		\label{fig:TimeUnboundMass} 
	\end{figure}

\begin{figure}
		\centering{}\includegraphics[scale=0.53]{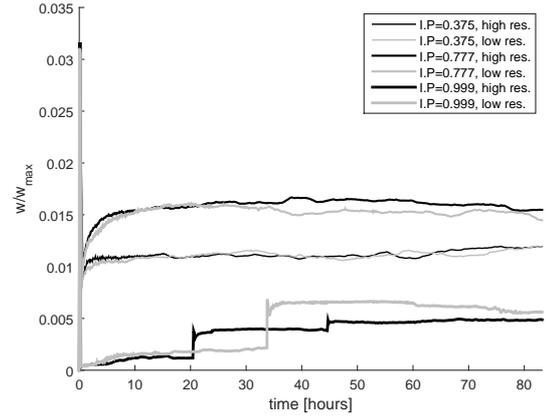} \caption{The proto-Earth's rotation rate as a function of time, given various impact parameters and SPH resolutions. For a 1/2 $M_{L}$ impactor mass.}
		\label{fig:TimeSpinRate} 
	\end{figure}

We explore the dependence of our results on the resolution of the SPH simulations, namely any potential difference between the resulting quantitites in similar simulations which differ in resolution. In particular we follow the evolution in time ($t$=0 marks the time of impact) of the disc mass, unbound mass and the proto-Earth's rotation rate. As mentioned earlier we compare our low resolution simulations ($5\cdot10^{4}$ particles) with the higher resolution ones ($10^{5}$ particles). The comparison is made for a non-rotating target (i.e., $w=0$). In Figures \ref{fig:TimeDiscMass}-\ref{fig:TimeSpinRate}, we show only simulations with an impactor mass of 1/2 $M_{L}$. Results for other masses are qualitatively similar.
	
	Figures \ref{fig:TimeDiscMass} and \ref{fig:TimeUnboundMass} show the bound debris disc mass and unbound material mass respectively, as a function of simulation time. Both are normalized with respect to the impactor's mass $M_{imp}$ (here = 1/2 $M_{L}$). Each figure contains several sub-plots, where line thickness correlates with impact parameter and line colour with SPH resolution. In order to account for the strong dependence on the choice of impact parameter, the y-axis is plotted in logarithmic scale, therefore zero mass points are not shown, leading to occasional discontinuous lines in Figure \ref{fig:TimeDiscMass}. 
	
	The differences between the low and high resolution simulations are generally small, as can be seen in Figure \ref{fig:TimeUnboundMass}, and differ only slightly in Figure \ref{fig:TimeDiscMass}, for extremely grazing impacts. In the latter case, there is a notable drop in the disc mass fraction occurring for the low resolution simulation after around $\sim$45 h, which does not exist in the high resolution simulation. The reason for this difference is that by the end of the simulation, one of the two largest fragments is unbound and the other one is classified as either in the disc (high resolution) or in the planet (low resolution). Since classification is based on the fragment pericentre relative to the fluid Roche limit, even subtle differences in fragment location as a result of resolution may lead to these changes in classification. Bound fragments often alternate their classification, which resembles a square wave function in the plot. As can be seen in Figure \ref{fig:TimeDiscMass} this mostly happens with small fragments, nevertheless, sometimes the difference is visually more noticeable when a large fragment is involved. Altogether, Figures \ref{fig:TimeUnboundMass}, \ref{fig:TimeDiscMass} and \ref{fig:TimeSpinRate} demonstrate a very close similarity. We thus consider performing the analysis on large sets of low resolution data justifiable, since the results will not be affected severely while providing significant improvement in computation expense.
	
	As one might expect, the normalized fractions of both bound and unbound masses correlate with the impact parameter. In other words, the more grazing the impact, the less mass accretes onto the planet and more remains in orbit around the target or completely escapes. 
	
	In Figure \ref{fig:TimeSpinRate} we show the change in the target's rotation rate as a result of the collision, from an initial non-rotating state ($w=0$). Since the infalling moonlet is not very massive in comparison to the target (here 1/2 $M_{L}$) the change in rotation rate is very small compared to the proto-Earth's breakup rotation rate $w_{max}$. As expected, the change is largest for intermediate impact angles, in which the largest amount of angular momentum is transferred to the target. In the grazing impact case, most of the change effectively arises from large fragments that rebound and re-collide with the planet at a later time. These can be seen as discrete jumps in the rotation rate, whereas otherwise the rotation rate changes more smoothly due to continuous interaction with small debris in the disc.
	
	\subsection{Spatial distribution of impactor material}\label{SS:Localization}
	Following the collision, the proto-Earth is mixed with material from the impacting moonlet. Due to the relatively low energies considered in this study, resulting from the typical collision velocity, and the small size of the impacting moonlet (and the smaller sizes of secondary collision fragments), we show that the mixing of material is far from homogeneous. This is in strike contrast to the typical outcomes of giant impact scenarios. The material is highly concentrated mostly around the initial impact site, and also in small amounts along the impactor-target collision plane (the equatorial plane of the target).
	
	In all the cases we have examined, the collision is insufficiently energetic in order to excavate core material from the target (see Section \ref{SS:Composition}). We thus have re-accreted silicate debris from the target and also silicate and iron from the impactor, noting that the assumed impactor iron core mass fraction is merely 2\% (see Section \ref{SS:ParameterSpace}). As noted above, in this section we wish to examine \emph{only} the distribution of impactor material, since the impactor is in itself the product of a previous collision with the proto-Earth, and as such its composition may differ from that of the proto-Earth. Non-uniformities in the distribution of impactor materials may thus support known isotopic heterogeneities in Earth silicates \citep{MarchiEtAl-2017}, as discussed in Section \ref{S:discussion}.
	
\begin{figure}
		\centering{}\includegraphics[scale=0.76]{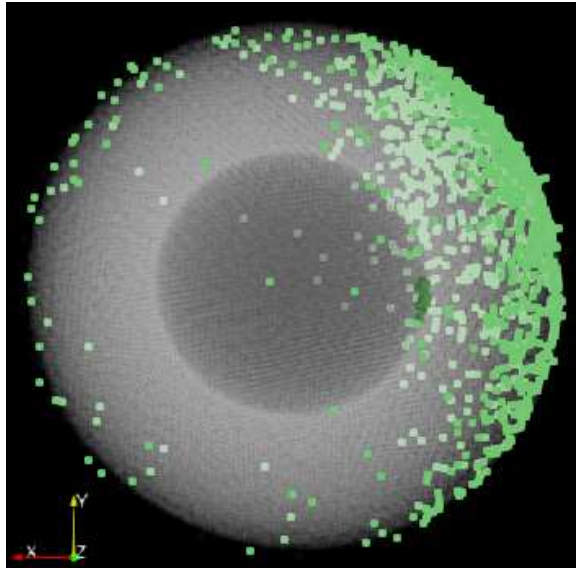} \caption{The distrbution of material in a high resolution ($10^{5}$ particles) simulation outcome for a 1/2 $M_{L}$ impactor and a 0.375 impact parameter, colliding with a non-rotating target. The image shows a semi-transparent, top view of the target. Iron (core) particles are denoted by dark pixels, while silicate (mantle) particles are light. Particles from the impactor are highlighted (green). We see that the distrbution of impactor particles is symmetric with respect to the xy collision plane, and that iron particles sink to the core of the target.}
		\label{fig:FullDistribution} 
	\end{figure}

\begin{figure*}
		\begin{centering}
			\subfigure[1/6 M$_L$ impactor - 0.375 impact parameter]{\label{fig:22_125}\includegraphics[scale=0.53]{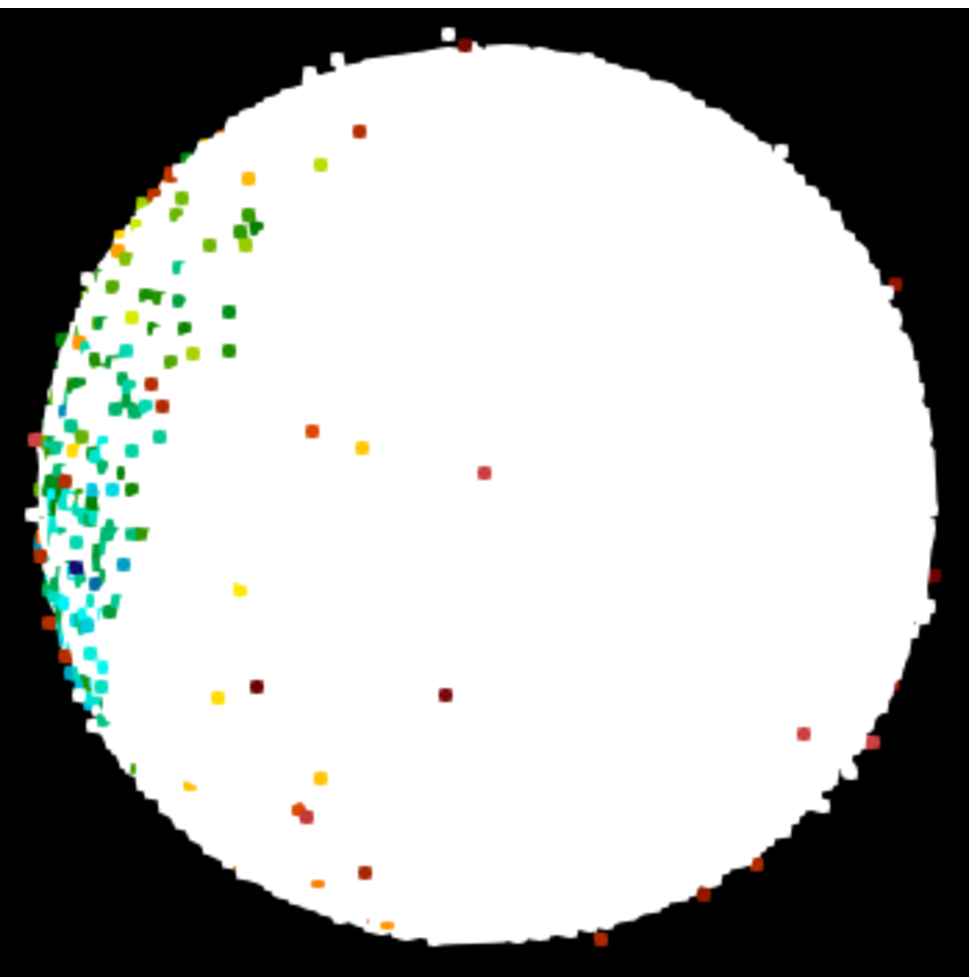}}
			\subfigure[1/6 M$_L$ impactor - 0.777 impact parameter]{\label{fig:51_125}\includegraphics[scale=0.53]{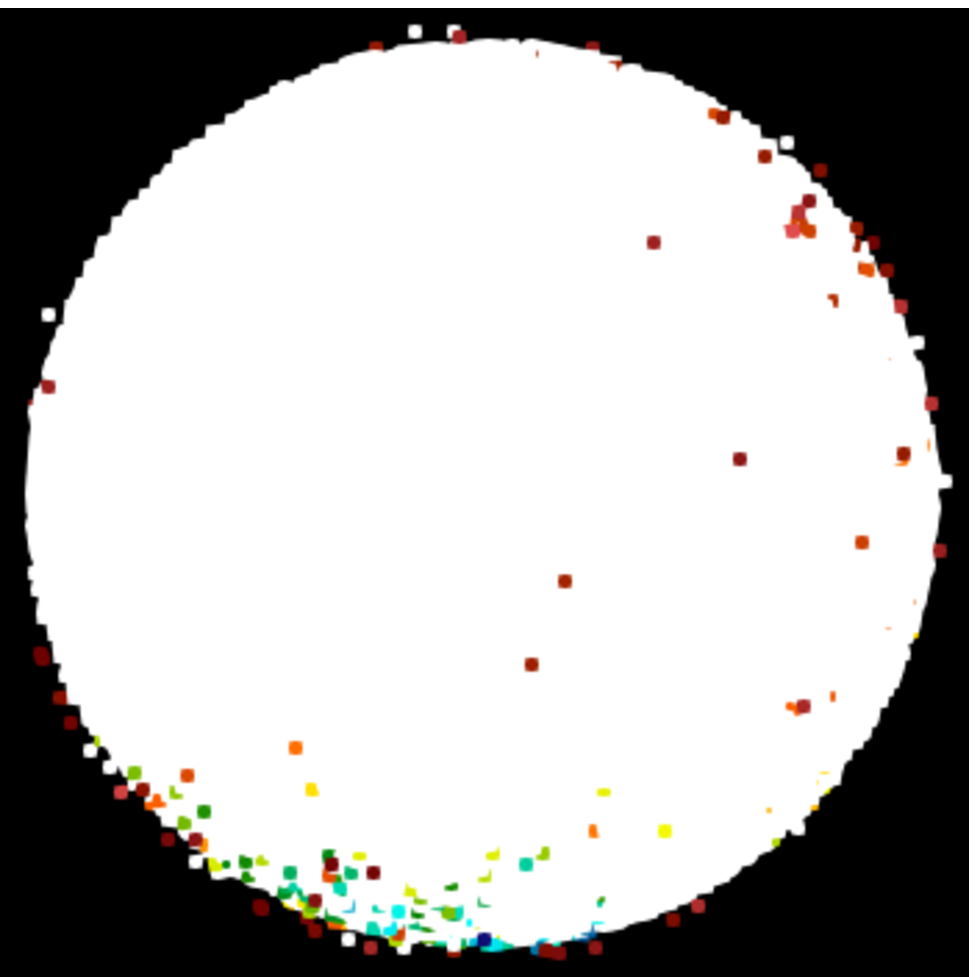}}
			\subfigure[1/6 M$_L$ impactor - 0.999 impact parameter]{\label{fig:89_125}\includegraphics[scale=0.53]{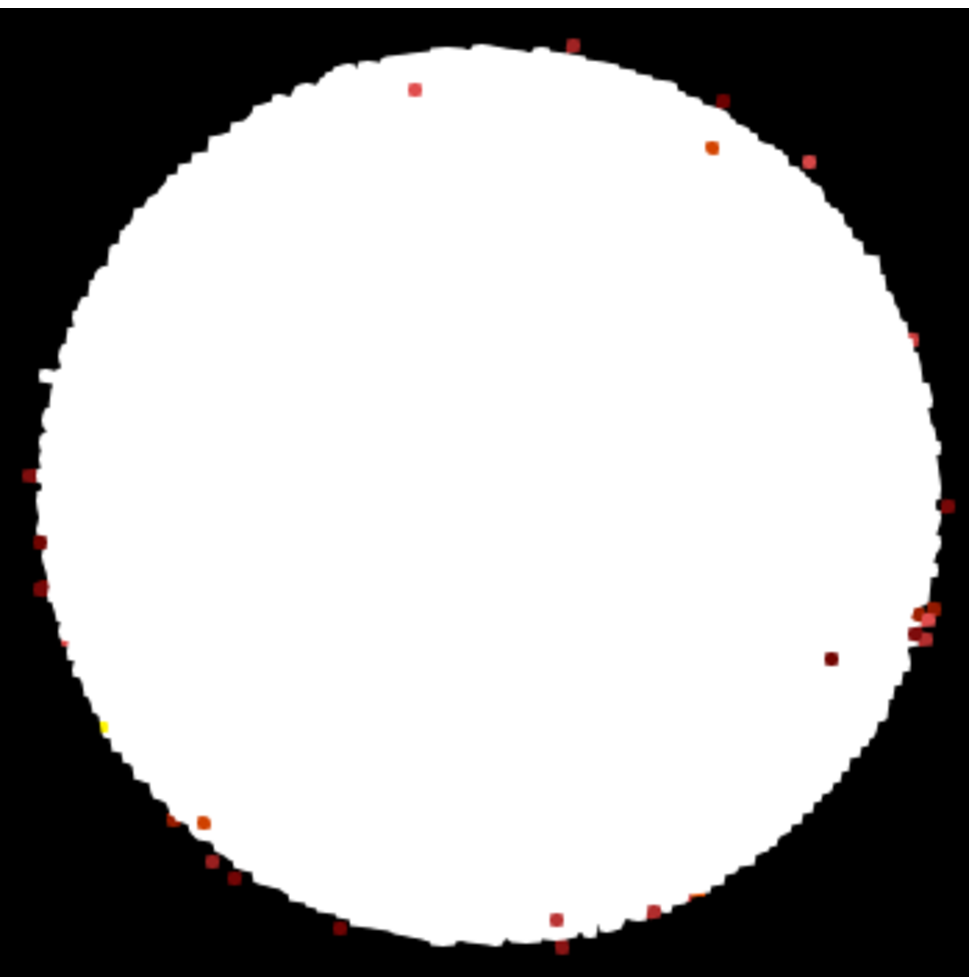}}
			\subfigure[1/3 M$_L$ impactor - 0.375 impact parameter]{\label{fig:22_25}\includegraphics[scale=0.53]{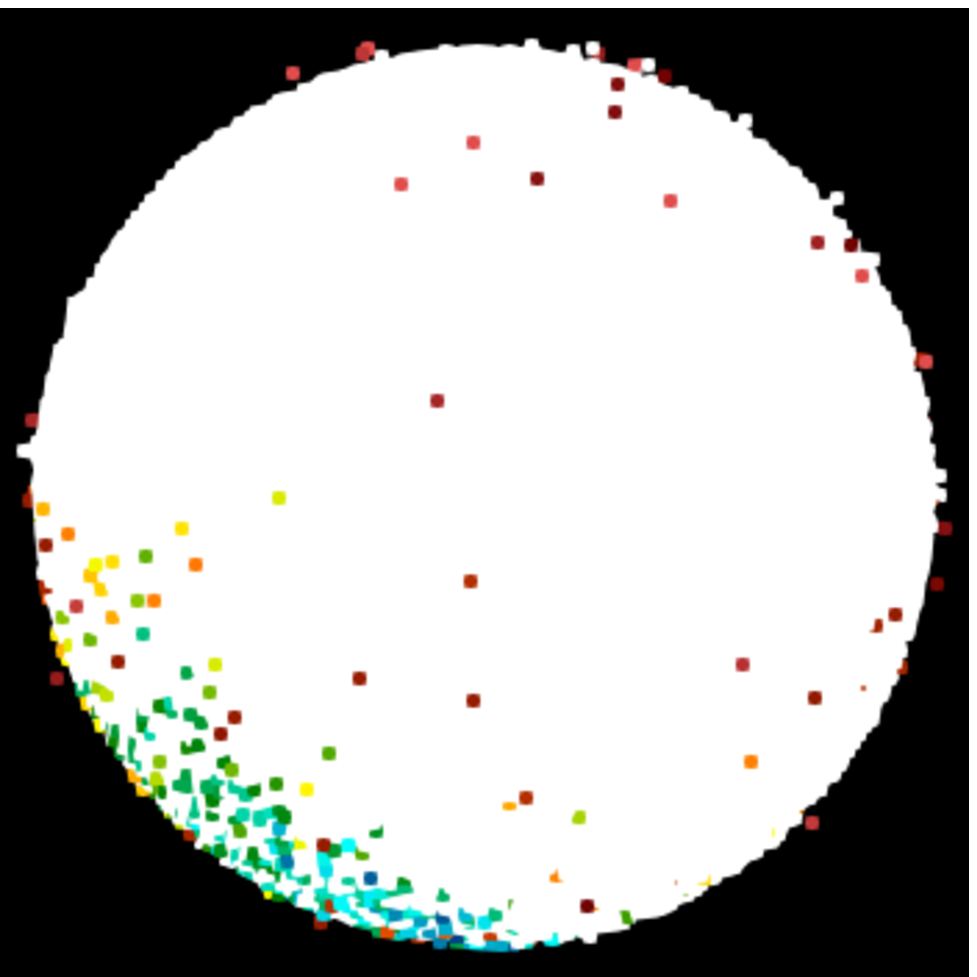}}
			\subfigure[1/3 M$_L$ impactor - 0.777 impact parameter]{\label{fig:51_25}\includegraphics[scale=0.53]{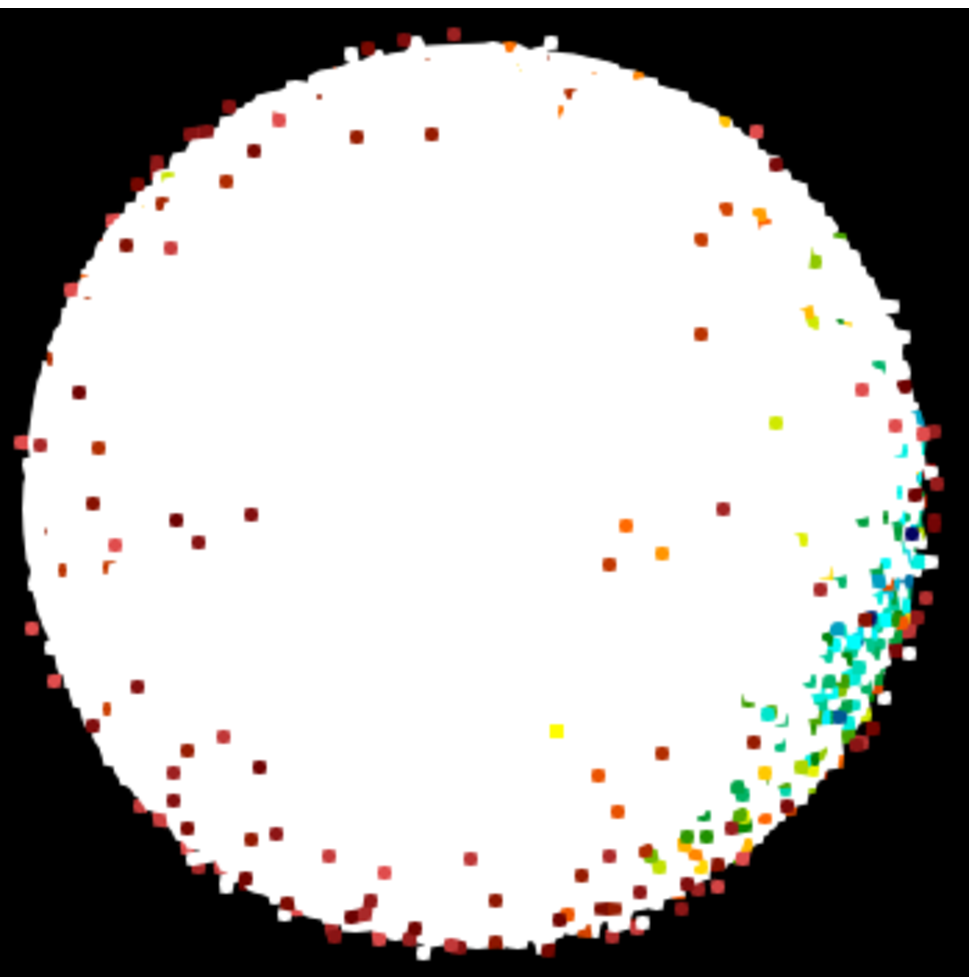}}
			\subfigure[1/3 M$_L$ impactor - 0.999 impact parameter]{\label{fig:89_25}\includegraphics[scale=0.53]{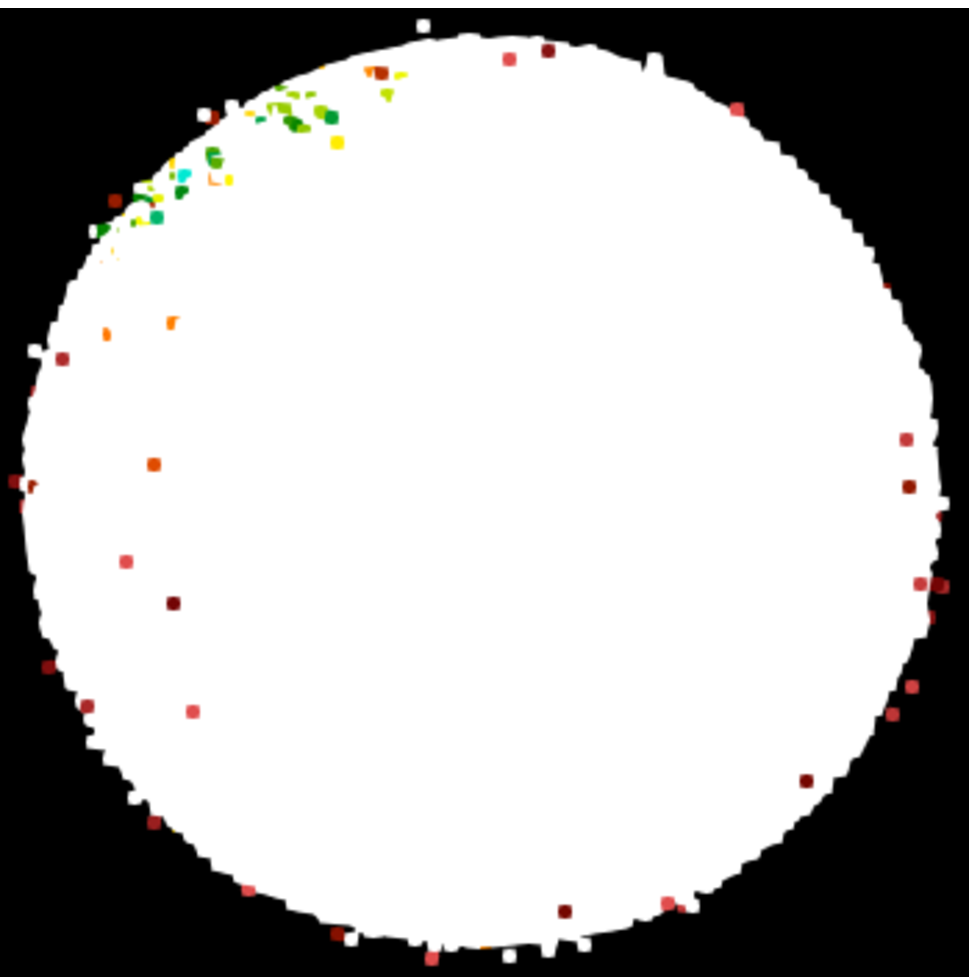}}
			\subfigure[1/2 M$_L$ impactor - 0.375 impact parameter]{\label{fig:22_37}\includegraphics[scale=0.53]{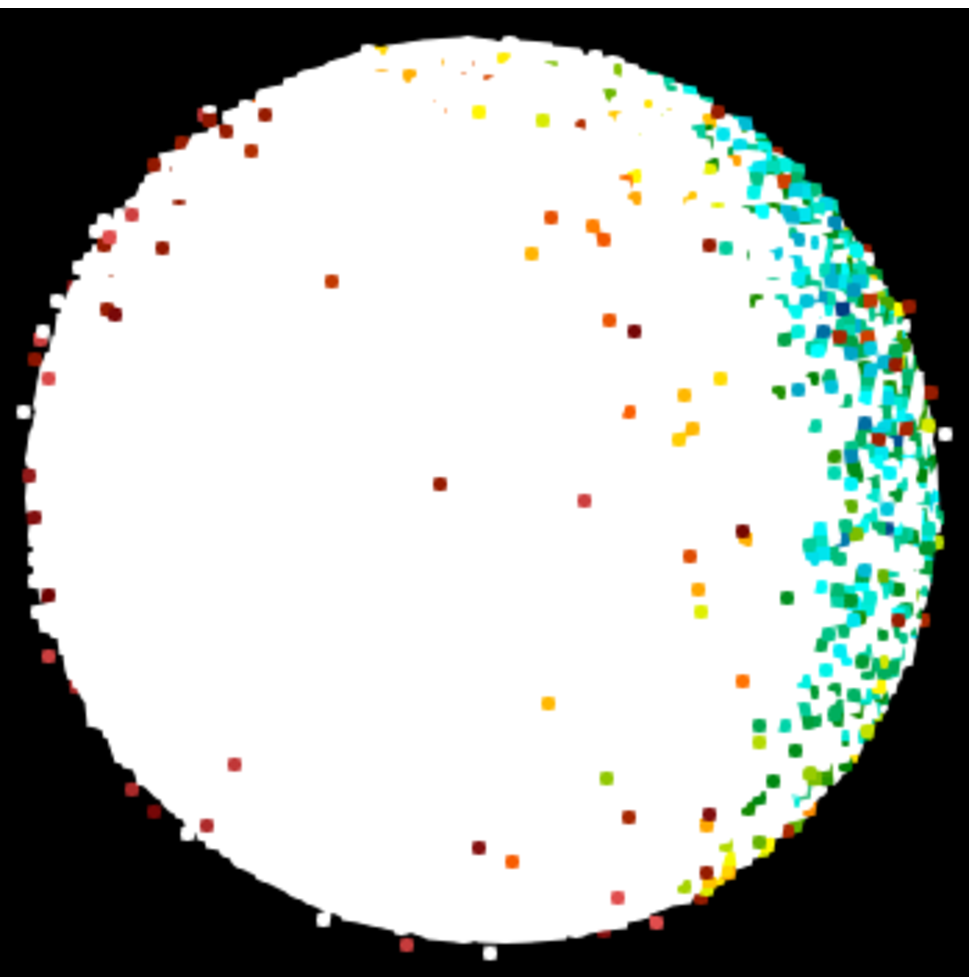}}
			\subfigure[1/2 M$_L$ impactor - 0.777 impact parameter]{\label{fig:51_37}\includegraphics[scale=0.53]{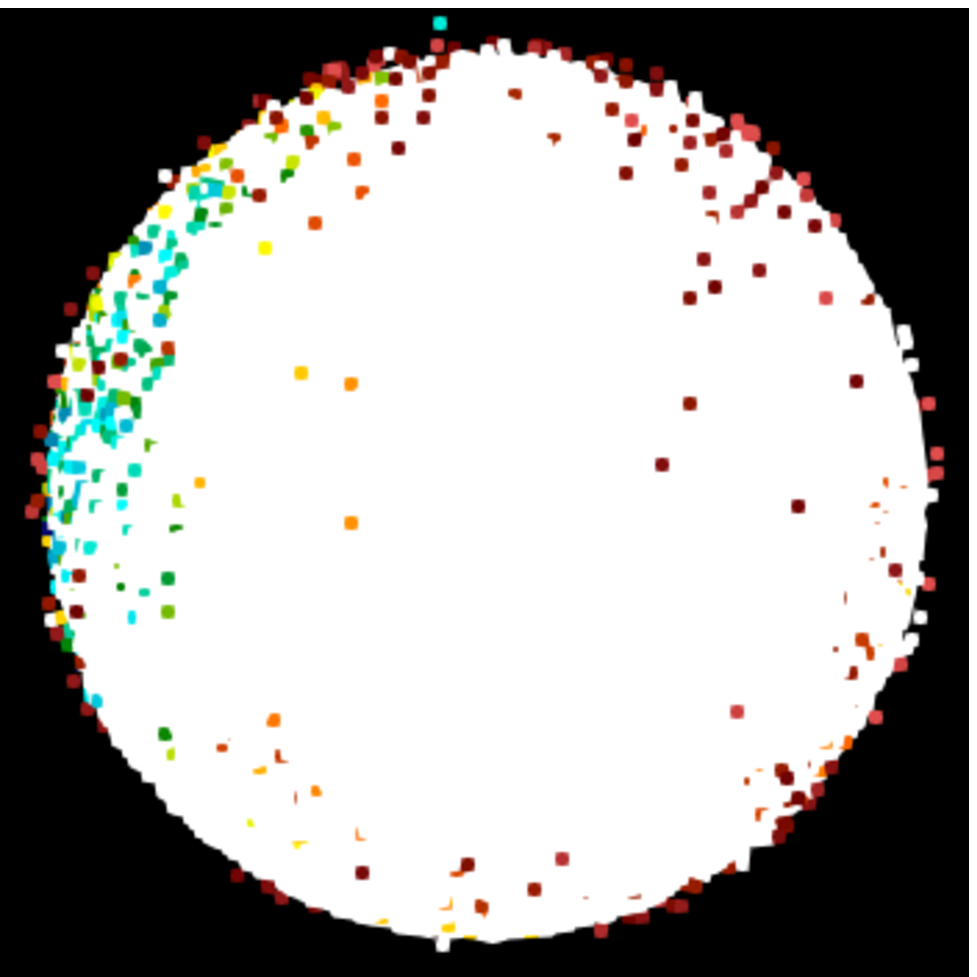}}
			\subfigure[1/2 M$_L$ impactor - 0.999 impact parameter]{\label{fig:89_37}\includegraphics[scale=0.53]{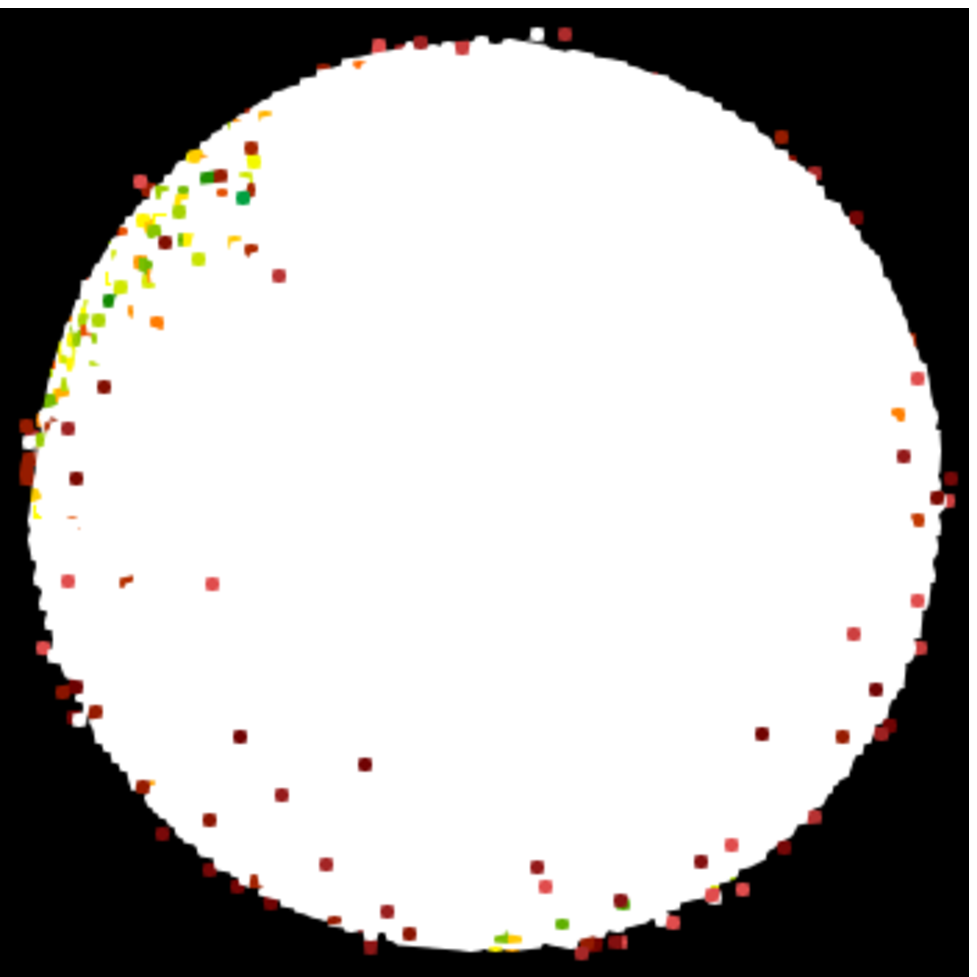}}
			\subfigure{\includegraphics[scale=0.55]{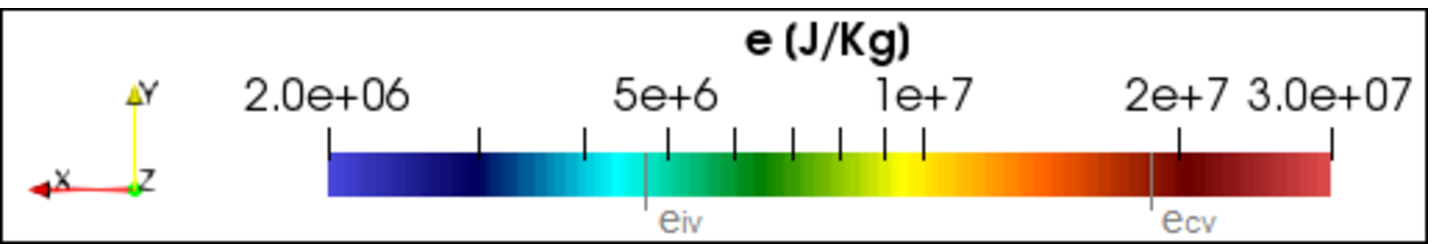}} 
			
		\end{centering}
		\caption{Panels (a-i) depict several high resolution ($10^{5}$particles) simulation outcomes for various impactor masses and impact parameters colliding with a non-rotating target. The images show an opaque, top view of the target. Impactor particles are coloured according to their specific internal energy $e$. Here accreted mass decreases with increasing impact parameter. The specific internal energy is typically close to or less than the energy of incipient vaporization.}
		\label{fig:OpaqueDistributions} 
	\end{figure*}

	Figure \ref{fig:FullDistribution} shows one example of a typical collision outcome at the end of the simulation. Here the impactor mass is 1/2 $M_{L}$, the collision is head-on into a non-rotating target and the simulation uses $10^{5}$ SPH particles. The figure shows a semi-transparent, top view (the xy plane is the collision plane) of the target, such that the darker iron core is clearly seen, underlying the lighter silicate mantle. Particles from the impactor are green-highlighted, where dark green pixels denote iron core material, green pixels denote silicate mantle material in the hemisphere facing the viewer and light green pixels denote silicate mantle material in the opposite hemisphere. As one might expect, from the viewer's vantage point we see that the impactor debris are quite symmetrically distributed with respect to the symmetry plane, lying on the equatorial plane of the target. In the subsequent Figure \ref{fig:OpaqueDistributions}
	we make use of this symmetry and plot only the observer-facing hemisphere without loosing much information. Note that the impactor iron core material sunk to settle on the target core-mantle boundary. In contrast, silicate material from the impactor settles on the surface of the target, as we further discuss below.
	
	Qualitative differences between the simulation presented in Figure \ref{fig:FullDistribution} and other simulation results arise when the mass of the impactor is smaller, or when the impact angle is higher. For the latter case, impactor iron core particles may end up in bound orbit around the target and/or even entirely unbound, instead of accreting on the target. Since the fraction of impactor iron particles is small in any case, it is convenient to focus on the distribution of impactor silicate material only, as we do in Figure \ref{fig:OpaqueDistributions}. Panels \ref{fig:22_125}-\ref{fig:89_37} show different distributions of the accreted impactor silicate material for simulations with various impactor masses and collision geometries. The transparent view is no longer required because of the aforementioned collision symmetry, and instead we provide a completely opaque, top view of the target, such that only impactor SPH particles that lie near or above the target surface are visible. As in Figure \ref{fig:FullDistribution}, all targets are initially non-rotating and the simulation uses $10^{5}$ SPH particles.
	
	Particle colours correspond to the specific internal energy $e$, and thus give us an approximation of their thermodynamic state. Note that near the collision contact point, the specific internal energy $e$ is often close to the energy of incipient vaporization $e_{iv}$ (4.72 $\cdot10^{6}$ J$\cdot$Kg$^{-1}$), or between that and the energy of complete vaporization $e_{cv}$ (1.82 $\cdot10^{7}$ J$\cdot$Kg$^{-1}$). Particles that are in the full vaporization regime are typically scarce and usually spread near the target's equatorial plane. While the Tillotson EOS used in this study provides merely a rough approximation of the thermodynamic state of the material, as discussed in Section \ref{SS:Outline}, we can qualitatively conclude that most of this accreted material is only partially molten/vaporized, and quite often it undergoes no phase transfer at all. Since the impactor is two orders of magnitude less massive than the target, its silicate mantle is expected to be less dense compared to that of the target to begin with. It is therefore not surprising to see this material settle on or near the surface of target, especially following an energetic impact. In some images we see a few particles that seem to hover above the surface. This is merely an indication of their high energy/low density, and since SPH particles all have the same mass \textendash{} the ones with lower density occupy a larger volume to this effect. The results for initially rotating targets are qualitatively identical to the results shown here.
	
	\subsection{Debris mass and composition}\label{SS:Composition}
\begin{figure*}
		\begin{center}
			\subfigure[Normalized disc mass - 0.375 impact parameter] {\label{fig:Md22}\includegraphics[scale=0.53]{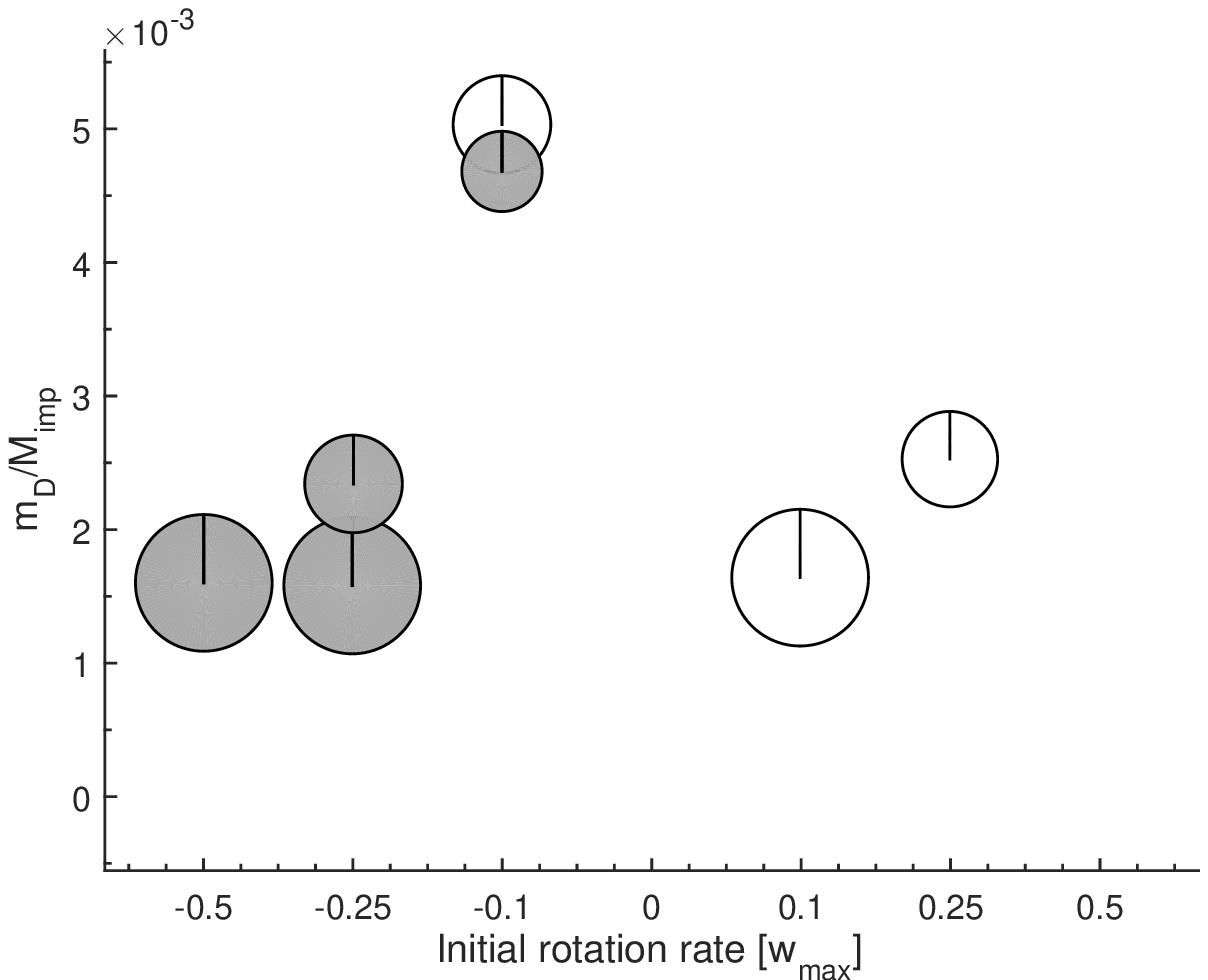}}
			\subfigure[Normalized unbound mass - 0.375 impact parameter] {\label{fig:Mu22}\includegraphics[scale=0.53]{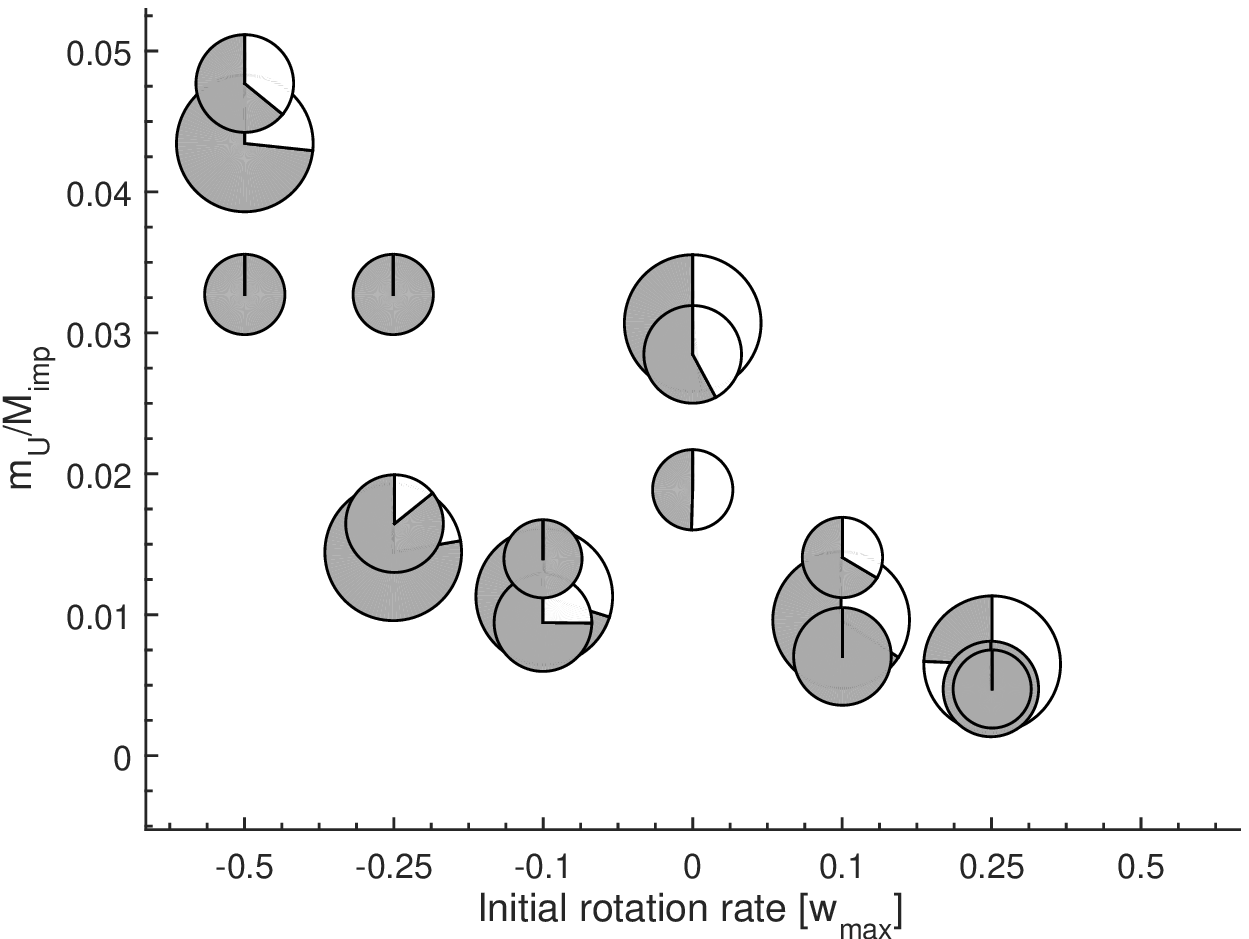}}
			\subfigure[Normalized disc mass - 0.777 impact parameter]
			{\label{fig:Md51}\includegraphics[scale=0.53]{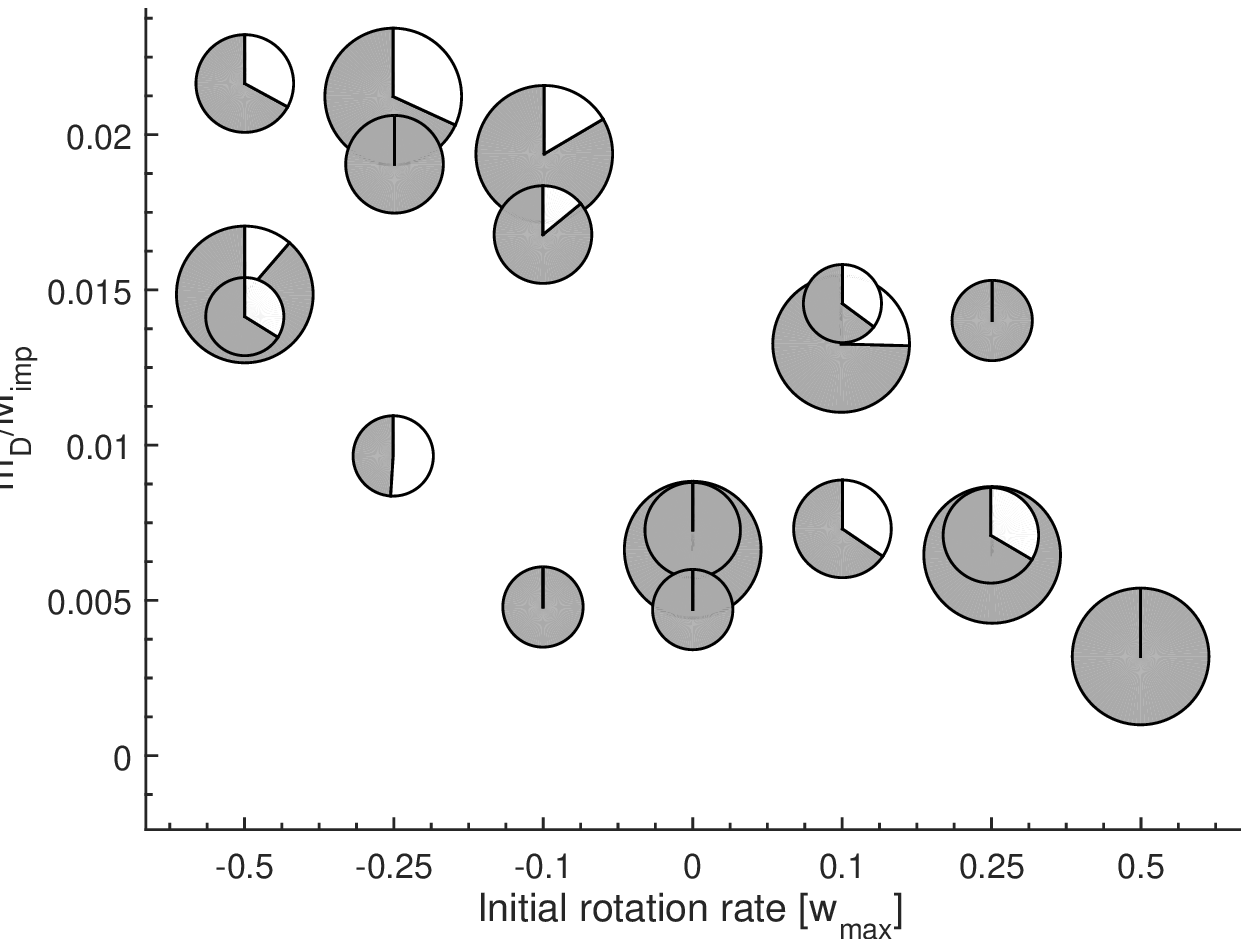}}		\subfigure[Normalized unbound mass - 0.777 impact parameter] {\label{fig:Mu51}\includegraphics[scale=0.53]{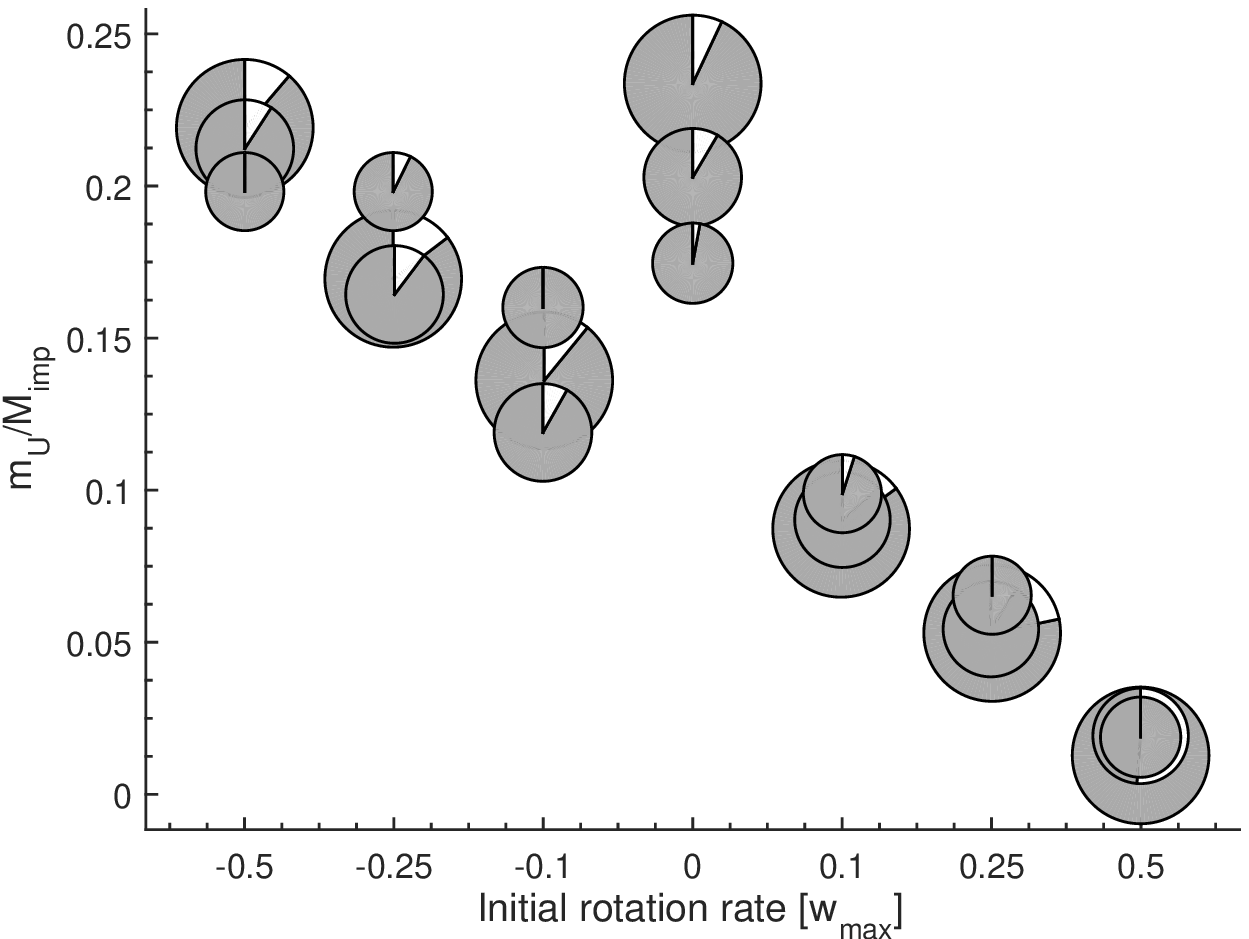}}
			\subfigure[Normalized disc mass - 0.999 impact parameter] {\label{fig:Md89}\includegraphics[scale=0.53]{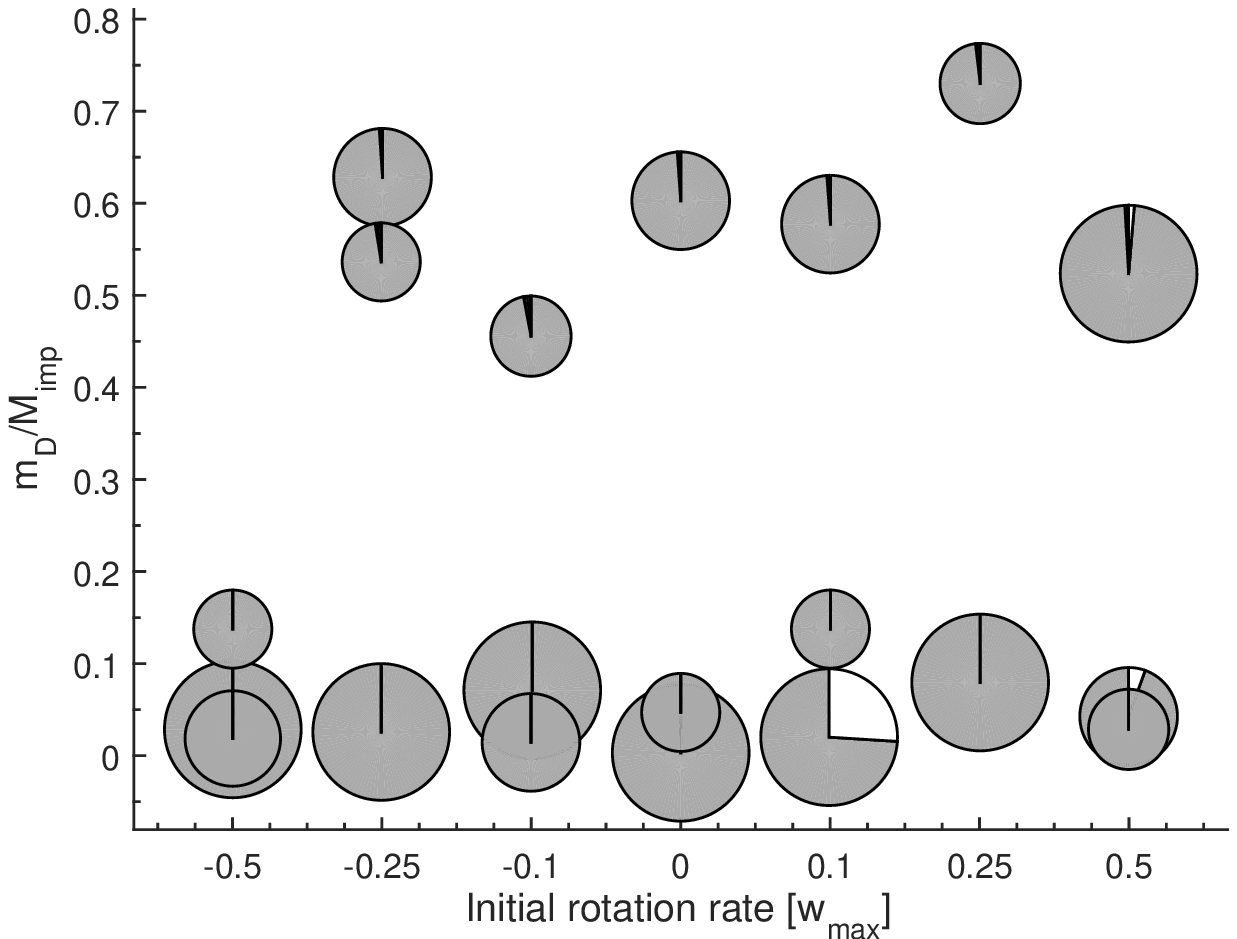}}		
			\subfigure[Normalized unbound mass - 0.999 impact parameter] {\label{fig:Mu89}\includegraphics[scale=0.53]{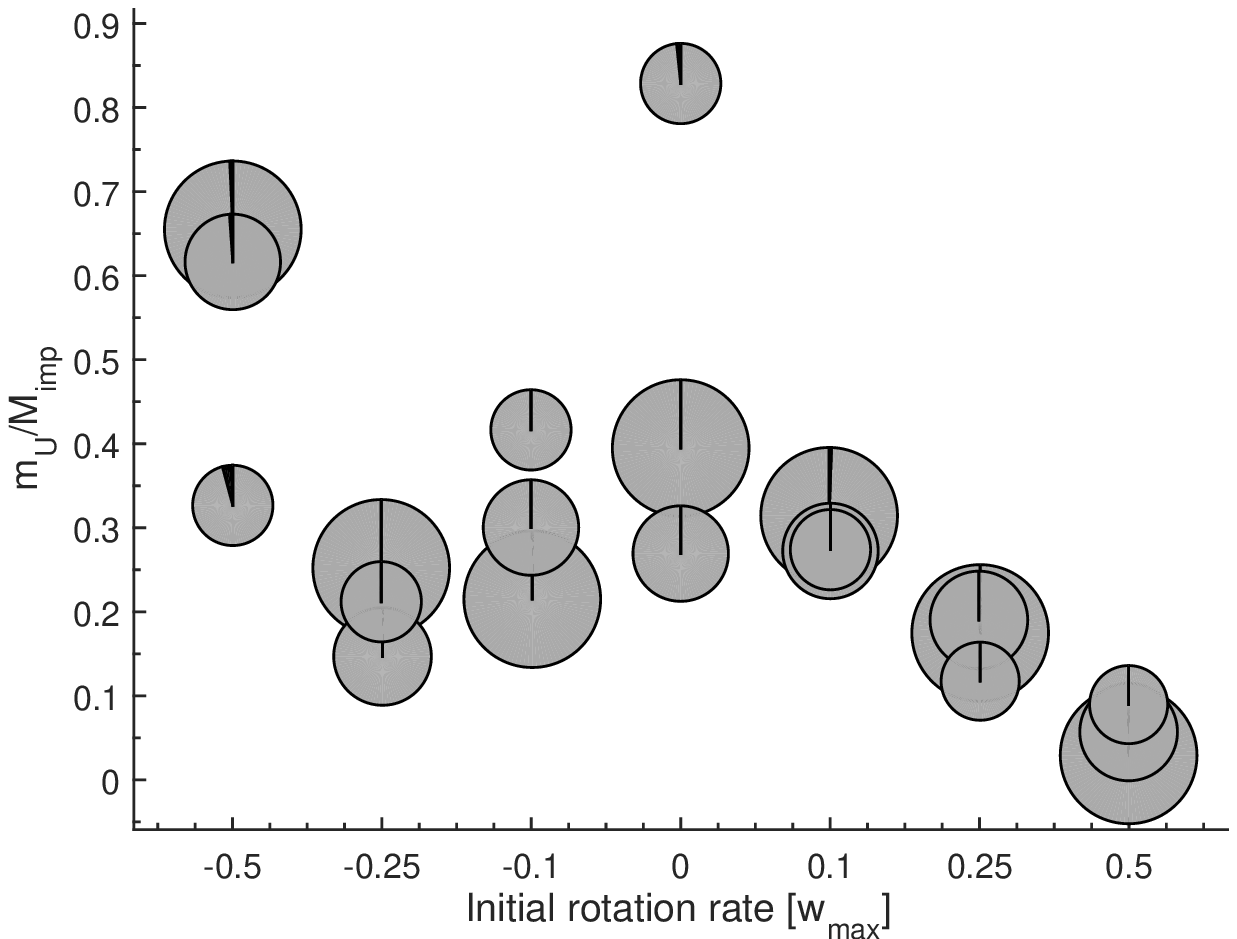}}
			\subfigure{\includegraphics[scale=0.43]{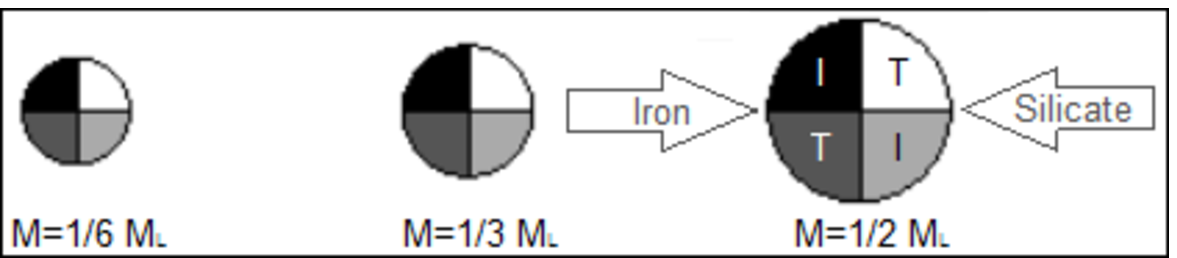}}
		\end{center}
		\caption{Debris mass and composition. The normalized disc mass is shown on the left column panels (a), (c) and (e), and the normalized unbound mass on the right column panels (b), (d) and (f), as a function of the target's initial rotation rate (negative signs denote retrograde collisions). Collision outcomes are presented by pie charts, colours showing the composition and size corresponding to the impactor mass. Pie charts are not drawn when the debris mass is zero.}
		\label{fig:Composition}
	\end{figure*}

	Figure \ref{fig:Composition} provides information about debris from the collision. The y-axis shows the normalized disc mass (with respect to the impactor mass) in the left column panels, and the normalized unbound mass in the right column panels, given different impact parameters. The x-axis shows the initial rotation rate of the target, where negative signs denote retrograde collisions. Each collision outcome is represented by a pie chart, where colours show the composition and size corresponds to the impactor mass as seen in the legend. The pie charts are missing in cases where the debris mass is zero.
	
	The results indicate that the debris are mainly composed of silicates from the impactor, and in some instances also from the target in smaller fractions. As we already mentioned in Section \ref{SS:Localization}, Iron \textendash{} if present \textendash{} only comes from the impactor and is never excavated from the target as it may be the case in more energetic giant impacts. For extremely grazing collisions all the impactor iron is typically found in the debris, rather than accretes onto the target. We also see that the mass of the debris is entirely negligible in near head-on collisions (thus a perfect merger is a good assumption), whereas it accumulates more significantly for higher impact parameters. Generally, there is up to an order of magnitude more mass in unbound material than there is in bound disc material, as well as a clear trend in the data, indicating more mass in retrograde collisions than there is in prograde collisions. 
	
	\subsection{Rotation rate}\label{SS:RotationRate}
\begin{figure*}	
	\begin{center}		
		\subfigure[The change in rotation rate $\Delta \omega$] {\label{fig:del_w}\includegraphics[scale=0.53]{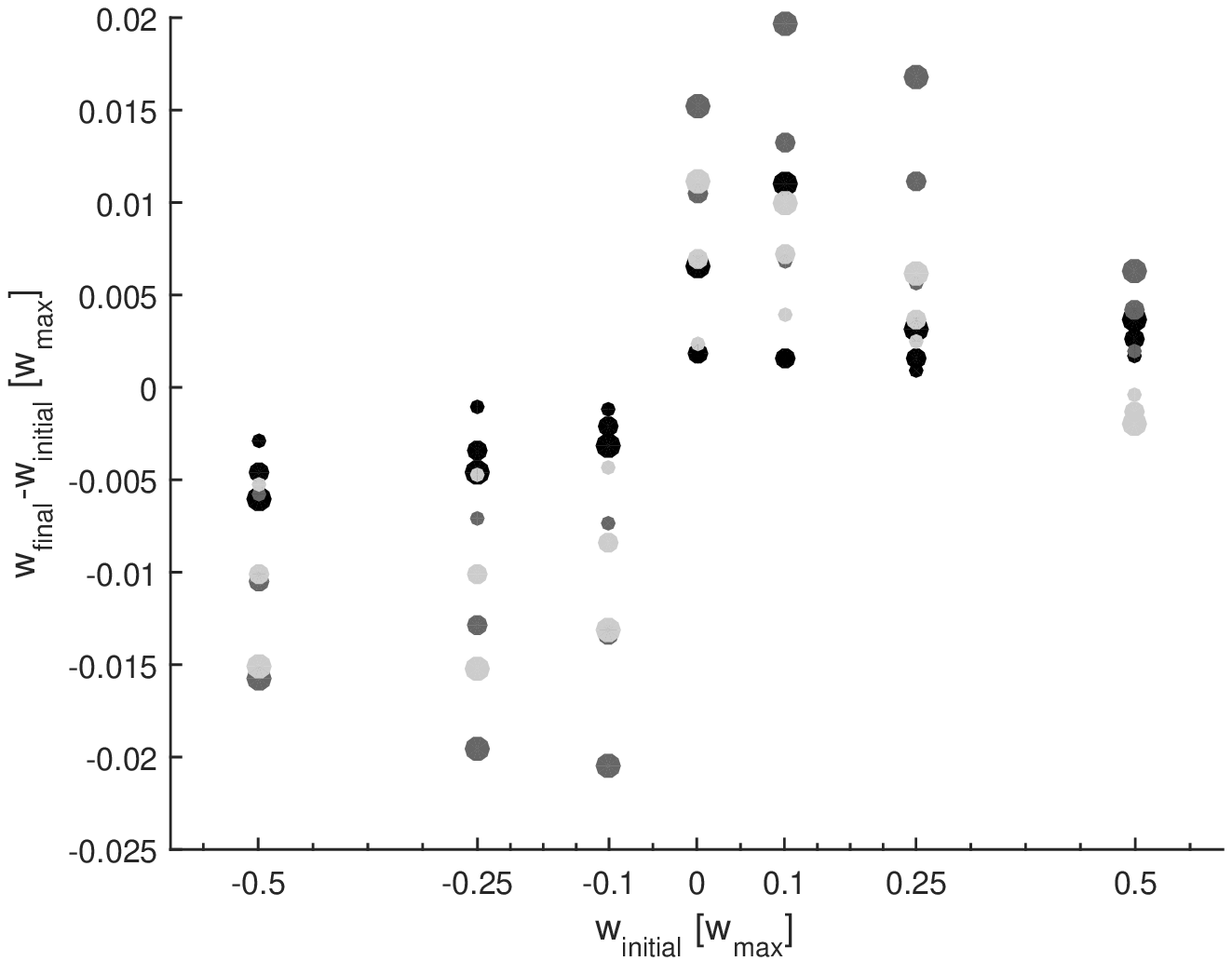}}
		\subfigure[The change in angular momentum $\Delta L$] {\label{fig:del_L}\includegraphics[scale=0.53]{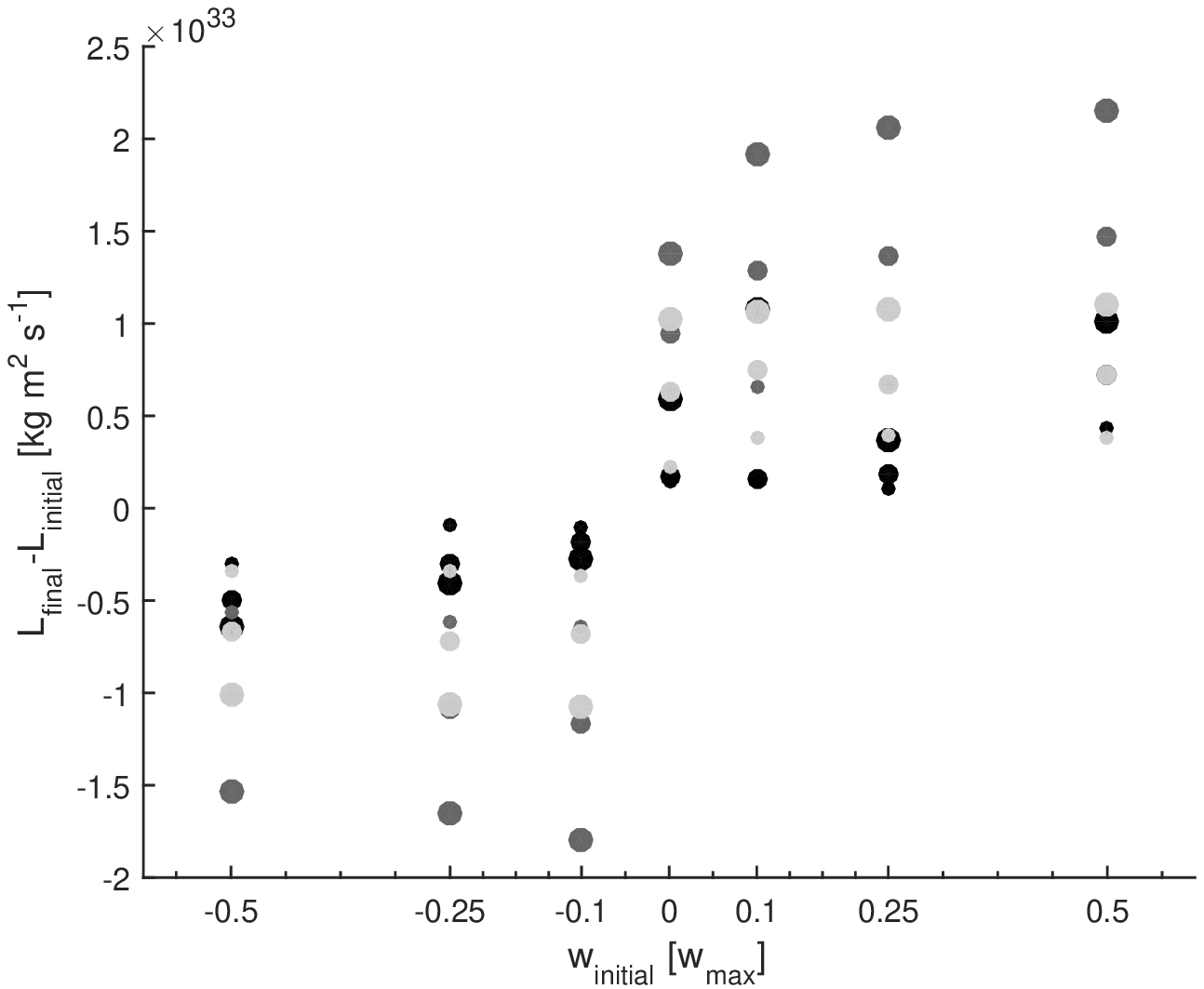}}
		\subfigure[The increase relative to the initial moment of inertia I]
		{\label{fig:inc_I}\includegraphics[scale=0.53]{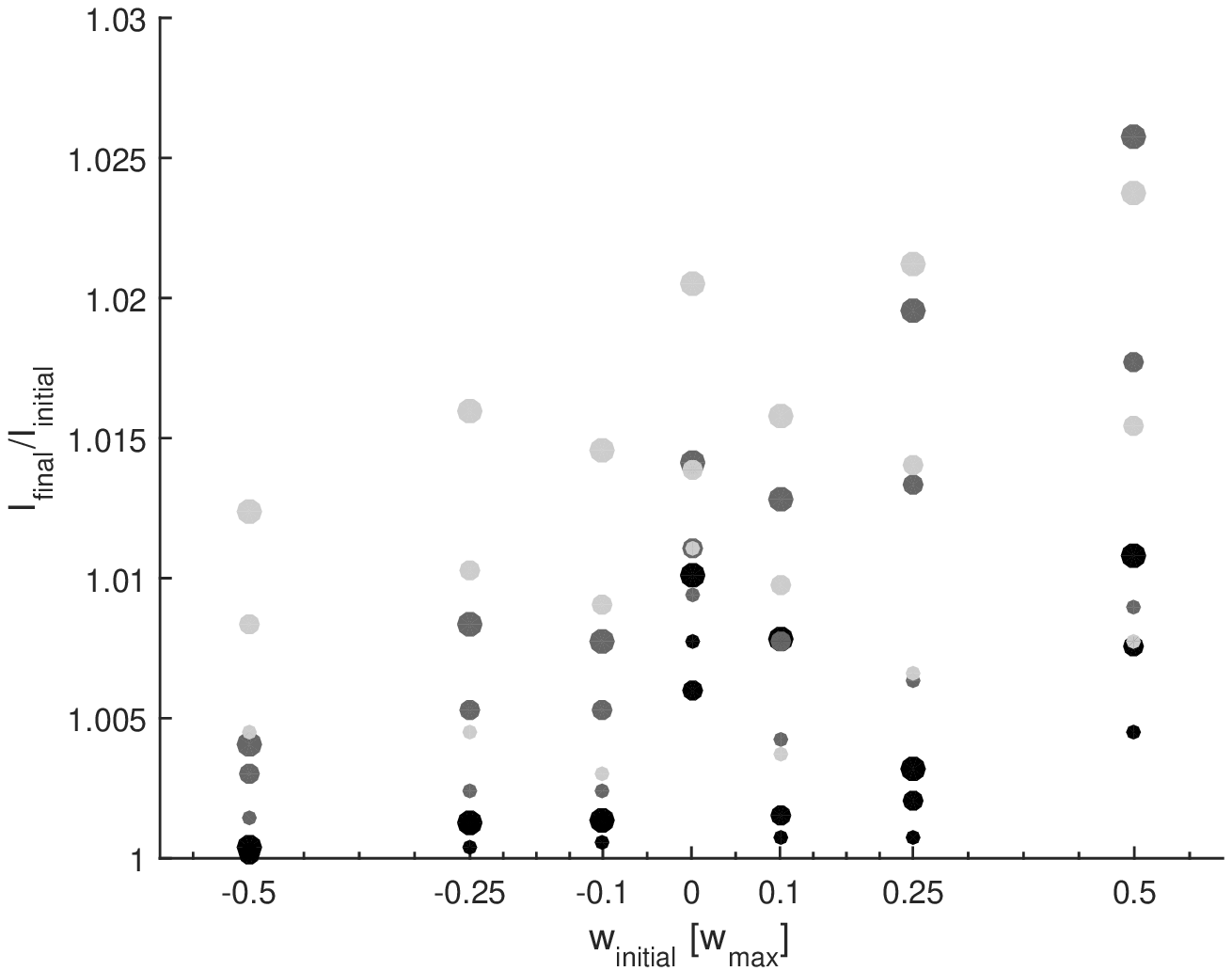}}
		\hspace{16.68em}
		\subfigure{\includegraphics[scale=0.65]{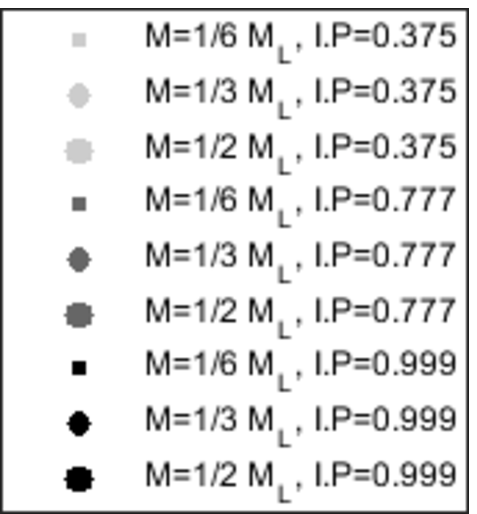}}
	\end{center}
	\caption{The change in the proto-Earth's rotation rate $\Delta \omega$ (a), angular momentum $\Delta L$ (b) and moment of inertia relative to initial value (c), as a function of the initial rotation rate (negative sign for retrograde collisions). Symbol sizes increase with impactor mass. Light grey symbols are for near head-on collisions, dark grey for intermediate collisions and black for extremely grazing collisions.}
	\label{fig:Rotation}
	\end{figure*}

	As a result of the collision, the initial rotation rate of the proto-Earth can change. As shown in Panel \ref{fig:del_w}, this change intricately depends on the impactor mass, the collision geometry and the exact initial rotation rate. The y-axis shows $\Delta\omega=\omega_{final}-\omega_{initial}$ and the x-axis the initial rotation rate (i.e., prior to the collision). Each collision outcome is represented by a filled circle, whose size increases with impactor mass, and whose colour corresponds to the impact parameter, black showing near head-on collisions, dark grey intermediate collisions and light grey extremely grazing collisions. As expected, retrograde collisions consistently decrease the target's rotation rate and $\Delta \omega$ is larger for more massive impactors. However, $\Delta \omega$ does not change equally for different initial rotation rates and also for prograde and retrograde collisions. This can be understood by examining the other two panels in Figure \ref{fig:Rotation}.  

	Panel \ref{fig:del_L} shows the change in the target's rotational angular momentum such that $\Delta L = L_{final}-L_{initial}$. As expected, $\Delta L$ is positive for prograde collisions and negative for retrograde collisions. However, for retrograde collisions $\Delta L$ is smaller when the initial rotation rate is higher and for prograde collisions the opposite is true. We explain this as a direct result of Figure \ref{fig:Composition}, which shows that the debris mass fraction is the largest for fast spinning retrograde collisions and smallest for fast spinning prograde collisions (see Section \ref{SS:Composition}). In other words, the more angular momentum carried by the debris \textendash{} the less rotational angular momentum remains in the target.
	
	Since $\omega=L/I$, we show the increase in the target's moment of inertia $I_{final}$ relative to its pre-collision value $I_{initial}$ in Panel \ref{fig:inc_I}. Here as well, when the debris mass fraction is larger the increase of $I$ is smaller, because the moment of inertia depends on the target's mass and radius and is therefore very sensitive to the amount of accreted mass. This leads to a dependence on the initial rotation rate, as in the previous paragraph, but also on the collision geometry. Near head-on collisions consistently produce negligible debris and thus increase $I$ more compared to other impact angles. The intricate interplay between $L$ and $I$ non-trivially produces the changes in $\Delta \omega$ seen in Panel \ref{fig:del_w}. The most prominent example of such intricate interplay is the outcome of a near head-on prograde collision onto a 0.5$\omega_{max}$ fast rotating target. Here $\Delta \omega$ actually decreases, i.e., the rotation rate slows down despite having been triggered by a prograde impactor. For this parameter combination the head-on collision increases $I$ the most and $L$ the least, to the extant that it is the only prograde case where $\omega$ decreases.
	
	We also find that changes in the direction of the rotation axis are entirely and consistently negligible. This outcome might have changed if the collision plane and the target's rotation axis were not assumed perpendicular. However such collision geometries remain to be investigated in future studies.
	
	\section{Discussion}\label{S:discussion}	
	The main objective of this paper is to examine the outcome of collisions between the proto-Earth (or similar-sized planets in general) and infalling moonlets. The paper is motivated from the need to understand the formation of the Earth-Moon system, in the larger context of the multiple impact origin hypothesis. In this hypothesis, several impacts during the late stages of planet formation form moonlets which gravitationally interact with the Earth and with previously formed moonlets, eventually giving rise to the formation of the Moon. One of several potential outcomes of this gravitational interaction is the orbital disruption of moonlets which causes them to re-collide with the proto-Earth. This particular aspect of moon-planet collisions, in characteristically low impact velocities, has never been studied in detail.
	
	We are interested in better quantifying specific aspects of the collision. Until now a common assumption in n-body simulations was to assume a perfect moon-planet merger. However our results in Section \ref{SS:Composition} indicate that such an assumption is perhaps only judicious for the more rare near head-on collisions. Otherwise for higher impact parameters an infalling moonlet could produce a considerable amount of debris. In extremely grazing collisions (and by extension we assume similar results would apply in tidal collisions) which are the most common to emerge from n-body simulations, the debris typically remain in large clumps of material, and are of comparable mass the original impactor. Bound clumps are essentially smaller moonlets that continue to evolve gravitationally around the proto-Earth. Therefore they may be viewed collectively as eroded versions of their progenitor, re-triggered to start a new evolutionary step resulting in a subsequent infall, merge or escape, and so forth.  
	
	In intermediate impact angles we find that formation of secondary moonlets in a bound debris disc could perhaps be neglected, since the disc mass does not exceed 2 \% of the impactor's mass. The fraction of impactor mass that could be ejected from the system however is substantial, and could be up to 25\%. We find that this fraction is independent of the impactor's mass, while at the same time highly dependent on the initial rotation rate of the target. It is highest for retrograde collisions onto fast rotating targets, and lowest for prograde collisions onto fast rotating targets. This trend repeats itself several times in our parameter space (Panels \ref{fig:Mu22}, \ref{fig:Md51}, \ref{fig:Mu51} and \ref{fig:Mu89}) \textendash{} the relative target and impactor velocities may either coincide resulting in collision dampening which lowers the amount of debris, or negate each other having the inverse effect.
	
	In cases where we find that the debris mass fraction is not negligible, we also find that the composition is typically very similar to that of the impactor, with very little material originating from the target. In other words, planetary erosion is found to be entirely and consistently negligible in moonfalls. Both of these results could certainly be applied to n-body simulations investigating the multiple-impact origin hypothesis, with the caveat that iron from the impactor \textendash{} if assumed \textendash{} should be omitted from the debris composition, except for extremely grazing or tidal collisions.
	
	The change in the proto-Earth's rotation rate as a result of the impact is found to have a complex dependence on various parameters. While in general retrograde collisions decrease the rotation rate and prograde collisions increase the rotation rate (with only one exception), the precise change is an intricate combination of the proto-Earth's angular momentum and moment of inertia. Both the former and the latter trivially depend on the mass of the impactor and the collision geometry, however they are also affected by angular momentum drain, carried away by debris from the impact which, as discussed in the previous paragraph, correlate with the magnitude and direction of the initial rotation rate. For initially slow rotating proto-Earth's the relative rotation rate change is the largest and could reach even 20-25\%. For initially very fast rotating proto-Earth's the relative rotation rate change is up to one order of magnitude smaller.
	
	Details resulting from this work can be used in order to approximate collisionally induced rotation rates in the proto-Earth (noting that collisions which are not in the Earth's equatorial plane are left for future studies), in addition to debris mass and composition. An additional useful result in this work comes from Section \ref{SS:Localization}, showing that the distribution of accreted impactor material on the proto-Earth is highly localized. Such heterogeneities are potentially important.
	
	In a recent paper \citet{MarchiEtAl-2017} re-evaluate the amount of mass that has been delivered to the Earth during its late accretion epoch (the so called 'late veneer')	using an SPH model for the Earth's bombardment by high velocity, relatively small differentiated planetesimals. An additional product of their work was the realization that known variations \citep{WillboldEtAl-2011,RizoEtAl-2016,DaleEtAl-2017} in highly siderophile elements incorporated in Earth mantle rocks, and in particular Tungsten ($^{128}$W), could be the natural outcome of such collisions, delivering impactor materials within discrete mantle domains. Tungsten is a decay product of its lithophile element precursor Hafnium ($^{128}$Hf) that has a short half-life of only 8.9 Myr, a fact that yields particularly strict constraints and makes it uniquely important. Here we argue, to complement their hypothesis, that collisions between the proto-Earth and low velocity infalling moonlets in the framework of the multiple impact origin, could essentially have the same effect, given our results in Section \ref{SS:Localization}. We note that a few infalling moonlets contribute a comparable amount of mass to late veneer estimations \citep{MarchiEtAl-2017}, and that unlike in the giant impact scenario, where at least partial if not full homogenization of the Earth's mantle is expected \citep{NakajimaStevenson-2015,PietEtAl-2017}, here the problem is easily circumvented given the smaller size of the impactors.
	 
	Finally, we note that the localization of debris accreted on the the proto-Earth is reminiscent of impacts on Mars and Charon, suggested to produce a hemispheric dichotomy \citep{WilhelmsSquyres-1984,Andrews-HannaEtAl-2008,MarinovaEtAl-2008,MalamudEtAl-2017}. Such low-energy impacts are sufficiently strong as to give rise to a large scale topographic structure on the impacted planets, but not sufficiently strong to give rise to planetary scale mixing. We therefore speculate that it is perhaps possible that a moonfall could have given rise to the first supercontinent structure on the Earth. Direct evidence for several past supercontinents go back only as far as 3.5 Gyrs, and these are thought to have formed through mantle convection processes (e.g. \citealt[and references therein]{Davies-1995,Condie-2004,ZhaoEtAl-2004}), however previous supercontinents may have existed before. In particular, a moonfall could have given rise to the first large scale planetary structure and the effective formation of an ex-situ formed supercontinent, which later evolved through in-situ geophysical processes that governed the evolution of subsequent supercontinents.
	
	\section{Summary}
	The terrestrial planets are thought to have formed during the first 10s-100s Myr of the evolution of the Solar-System. During the last stages of terrestrial planet formation the planets grew mainly through impacts with large planetary embryos. According to the current paradigm for the origins of the Moon, it is thought to have formed during this stage, following the last giant-impact with the proto-Earth. In this scenario the Moon coagulated from the debris disk that formed following this impact. Since then the Moon has evolved due to the mutual Earth-Moon tides, and has migrated outwards to its current location. However, the current paradigm is intrinsically incomplete and disconnected from the wider picture of terrestrial planet formation, in which the proto-Earth had experienced and grown through multiple giant impacts. Consideration of only the last giant impact in the current paradigm disregards the critical evolution taking place prior to (and possibly following) this event. Recently we proposed to explore a novel multiple-impact scenario, which is naturally aligned with the global context of planet formation. In this scenario multiple moons have formed around the proto-Earth following multiple planetary-scale impacts in the past, a sequence that has thereby affected the evolution of both the Earth and the Moon. 
	
	In our novel scenario the moon formation process described above occurs repeatedly after each planetary-scale impact on the proto-Earth. Therefore a moonlet may already exist when another impact forms a new debris disk and eventually, another moonlet. The new moonlet forms closer to the planet than the previous one(s) which already migrated outwards. Since tidal evolution is faster closer to the planet the new moonlet may migrate outwards and catch-up with the previous one on a typical timescale of Myr. Once the two moons closely approach they strongly interact through gravitational chaotic evolution. Such interactions could lead to moon-moon collisions, to moon collisions with the Earth (moonfalls) or to ejection of moons from the system. Retrograde planetary-scale collisions may also give rise to retrograde de-orbiting moons that will ultimately collide with the Earth in the same way. The overall multiple-moon formation and dynamics could therefore play a key role in determining the orbital and physical properties of the Earth-Moon system, both before and after the formation of the current Moon.
	
	In this paper we explore the outcomes of moonfalls, namely the cases were past Earth moonlets collided with the Earth. We make use of a SPH collision model that utilizes graphic processing units for highly improved computing time performance and consider a wide range of possible collsion configurations. Our main findings are as follows:
	\begin{itemize}
		\item Grazing or tidal collisions are the most typical type of impacts and they produce debris (both bound and unbound) whose mass-fraction is comparable to that of the original impactor. Most of the debris typically remain in large clumps of material. Bound clumps may therefore be viewed as small moonlets \textendash{} eroded version of their progenitor. Each small moonlet continues to gravitationally evolve around the proto-Earth, either de-orbiting or tidally evolving outward and interacting with a pre-existing/later-formed moonlet.
		\item Intermediate impact angles result in debris mass fractions in the range of 2-25\% where most of the material is unbound. 	\item Retrograde collisions produce more debris than prograde collisions, and the exact mass fraction depends on the proto-Earth initial rotation rate. 
		\item The collision-induced change in the initial rotation rate is generally small, consistent with the small amount of angular momentum brought by low-mass moonlets.
		\item The material accreted on the proto-Earth from an infalling moonlet is highly localized, potentially explaining the isotopic heterogeneities in highly siderophile elements in terrestrial rocks. We speculate that the localized mass could have given rise to the first primordial super-continent topographic features. 
	\end{itemize}
	We note that these results can be used for simple approximations and scaling laws and applied to n-body/Monte-Carlo studies of the formation of the Earth and Moon through a series of multiple impacts.

	\section{Acknowledgment}\label{S:Acknowledgment}
	UM and HBP acknowledge support from the Marie Curie FP7 career integration grant "GRAND", the Research Cooperation Lower Saxony-Israel Niedersachsisches Vorab fund, the Minerva center for life under extreme planetary conditions, the Israeli Science and Technology ministry Ilan Ramon grant and the ISF I-CORE grant 1829/12. CS and CB acknowledge support by the high performance and cloud computing group at the Zentrum f{\"u}r Datenverarbeitung of the University of T{\"u}bingen, the state of Baden-W{\"u}rttemberg through bwHP and the German Research Foundation (DFG) through grant no INST 37/935-1 FUGG. CB acknowledges support by the FWF Austrian Science Fund project S11603-N16.

	
	
	\bibliographystyle{mnras} 
	\bibliography{bibfile}     
	
	\bsp	
	\label{lastpage}
\end{document}